\documentclass[twocolumn]{aastex631}

\usepackage{appendix}

\usepackage{amsmath}
\usepackage{comment}
\received{January 21, 2025}
\accepted{April 18, 2025}
\submitjournal{ApJ}

\def\gtrsim{\lower 2pt \hbox{$\, \buildrel {\scriptstyle >}\over
{\scriptstyle \sim}\,$}}
\def\lesssim{\lower 2pt \hbox{$\, \buildrel {\scriptstyle <}\over
{\scriptstyle \sim}\,$}}

\def\xmm{{\sl XMM-Newton}}

\def\chandra{{\sl Chandra}}

\newcommand{\as}{$^{\prime\prime}$}

\shorttitle{}

\graphicspath{{./}{figures/}}

\begin{document}
\title{Diffuse X-ray emission in M51: a hierarchical Bayesian spatially-resolved spectral analysis}
\author[0009-0008-2940-6166]{Luan Luan}\thanks{Contact e-mail: luanluan@umass.edu} 
\author[0000-0002-9279-4041]{Q. Daniel Wang} \thanks{Contact e-mail: wqd@umass.edu} 
\affiliation{Department of Astronomy,University of Massachusetts, Amherst MA 01003-9305, USA}

\begin{abstract}
X-ray observations can be used to effectively probe the galactic ecosystem, particularly its hot and energetic components. However, existing X-ray studies of nearby star-forming galaxies are limited by insufficient data statistics and a lack of suitable spectral modeling to account for X-ray emission and absorption geometry. We present results from an X-ray spectral study of M51 using 1.3-Ms Chandra data, the most extensive for such a galaxy. This allows the extraction of diffuse X-ray emission spectra from spiral arm phase-dependent regions using a logarithmic spiral coordinate system. A hierarchical Bayesian approach analyzes these spectra, testing models from simple 1-T hot plasma to those including distributed hot plasma and X-ray-absorbing cool gas. We recommend a model fitting the spectra well, featuring a galactic corona with a lognormal temperature distribution and a disk with mixed X-ray emissions and absorption. In this model, only half of the coronal emission is subject to internal absorption. The best-fit absorbing gas column density is roughly twice that inferred from optical extinction of stellar light. The temperature distribution shows a mean temperature of $\sim 0.1$ keV and an average one-dex dispersion that is enhanced on the spiral arms. The corona's radiative cooling might balance the mechanical energy input from stellar feedback. These results highlight the effectiveness of X-ray mapping of the corona and cool gas in spiral galaxies.  

\end{abstract}

\section{Introduction} \label{sec:intro}

 X-ray emission from active star-forming spiral galaxies generally tracks the end products of stars \citep[e.g.,][]{Fabbiano1992,Wang2016}.
The brightest X-ray sources are typically powered by accretion: active galactic nuclei and low-mass or high-mass X-ray binaries (HMXBs). Such objects with X-ray luminosities of $\gtrsim 10^{37} {\rm~ergs~s^{-1}}$ can be detected individually in nearby galaxies ($\lesssim 10$~Mpc) with a reasonably deep \chandra\ exposure (e.g., $\gtrsim 10^2$~ks) and can be largely excised from the data.  The remaining so-called diffuse X-ray emission mostly represents the hot components of the interstellar medium (ISM) and the circumgalactic medium (CGM). Significant contamination from numerous fainter discrete sources (including cataclysmic variables and active binaries) can be estimated and statistically subtracted in the data analysis. The total luminosity of the diffuse X-ray emission is well correlated with the star formation rate and stellar mass of galaxies and is thought to trace the feedback from active galactic nuclei, as well as from stars, both young and old, in the form of stellar winds and supernovae \citep[e.g.,][]{Li2013,Zeng2023}. 

Studies of diffuse X-ray emission have largely focused on highly inclined disk galaxies. The high inclination allows observations of the extraplanar hot plasma or galactic corona, which typically emits soft X-rays, with minimal contamination from discrete sources that are predominantly located in galactic disks \citep[e.g.,][]{Strickland2004,Li2013}. However, soft diffuse X-ray emission typically accounts for only about 1\% of the energy input rate of stellar feedback \citep[see][]{Mineo2012,Li2013}.  It is not clear whether most of the energy input is consumed within the galactic disks (including their immediate vicinity) or escapes in other ways (e.g., galactic winds or outflows). The connection between the extraplanar hot plasma and the star-formation activity in the galactic disks cannot be easily probed because of the line-of-sight obscuration and confusion in such galaxies. 

Progress in understanding the diffuse X-ray emission in face-on-disk galaxies has also been slow. Although the emission is relatively easy to detect and shows a clear spatial correlation with spiral arms of a galaxy \citep[e.g.,][]{Tyler2004}, it has been difficult to infer the properties of the hot plasma due to several complications. First, internal X-ray absorption is challenging to model because it primarily occurs in the galactic disk. The usual foreground modeling of the absorption is clearly problematic. Second, the absorption may vary strongly from region to region (e.g., from on-arm to off-arm regions) and on all scales. Third, the plasma distributed from the galactic disk to the corona is expected to have a range of properties, e.g., different temperatures.  These effects have been demonstrated in a recent study of the face-on galaxy M83 \citep{Wang2021}, using very deep \chandra\ observations.  It is found that the spectra of the diffuse X-ray emission are better characterized by a lognormal temperature distribution model plus an inhomogeneous absorbing gas distribution. So far, such a study is based on the independent analysis of the spectra extracted from different regions (e.g., different spiral arms), which may represent different realizations of the same underlying astrophysical processes. The quality of individual spectra remains too limited to map the distribution of plasma properties. Ideally, they should be analyzed together to improve the constraints on the model parameters and their variation with the astrophysical conditions.  

Here we report our study of the nearby face-on grand spiral Sa galaxy M51  \citep[NGC\,5194, the Whirlpool Galaxy; Table~\ref{t:M51}; e.g., ][]{2019MNRAS.488..590B}  to test a new approach to diffuse X-ray spectral analysis. The galaxy is ideally suited for this test for the following reasons. First, it has been extensively observed with \chandra\, accumulating a total of 1.3 Ms of exposure -- the longest ever for a nearby galaxy. Second, the relative proximity and small inclination angle of the galaxy make it easier to identify spiral arms versus inter-arm regions. Third, the low Galactic foreground H\texttt{I} column density maximizes the sensitivity to soft X-ray emission expected from the diffuse hot plasma or the galactic corona of the galaxy.  The new spatially resolved spectral analysis method, based on the hierarchical Bayesian approach, allows us to jointly fit \chandra\ spectra extracted from regions at different spiral arm phases with an optimal number of fitting parameters and to characterize the mean and variation of key spectral parameters such as plasma temperature and X-ray-absorbing gas column density.  As a result, we are able to test more and more physical spectral models to explore the temperature and absorption distributions and their dependence on the arm phase.

The remainder of the paper is organized as follows. In Section \ref{sec:data}, we describe the data reduction and analysis methodology. Section \ref{sec:result} presents various models analyzed and the results of fitting these models. In Section \ref{sec:discussion}, we discuss the implications of the results for the properties of hot plasma and stellar feedback. Section \ref{sec:summary} summarizes our results and conclusions. All error bars are quoted at the confidence level of 1 $ \sigma $.
\begin{deluxetable}{lcr}
\tablenum{1}
\tablecaption{Adopted Parameters of M51}
\tablewidth{0pt}
\tablehead{
\colhead{Parameter} & \colhead{Value}  & \colhead{reference} \\
}
\startdata
R.A. (J2000)  & $13^\circ29^{\prime}52.771^{\prime\prime}$ & \cite{Pineda2020}\\
Decl. (J2000)  & $+47^\circ11^{\prime}42.62^{\prime\prime}$ & \cite{Pineda2020} \\
Pitch angle ($i_{pitch}$)\tablenotemark{a} & $-18.5^\circ$ & \cite{Pineda2020} \\
Position angle ($\theta_{PA}$) & $20.2^\circ$ & \cite{Hu2013} \\
Inclination angle ($i_{inc}$) & $12.0^\circ$ & \cite{Hu2013} \\
Distance  & 8.4 Mpc & \cite{Shetty2007} \\
Galactic H\texttt{I} column  & $3.3 \times 10^{20} {\rm~cm^{-2}}$ & \tablenotemark{b}\\
\hline
\enddata
\tablenotetext{a}{Applicable to the annulus region studied in this work with $r_{min}=45$\as\  and $r_{max}=110$\as.}
\tablenotetext{b}{\cite{HI4PI2016}.}
\label{t:M51}
\end{deluxetable}

\section{Data Reduction and Analysis} \label{sec:data}

Table \ref{t:observations} lists the \chandra\ observations used in the present work.
The pointing positions of these observations are all within the 3$^{\prime}$ off-axis radius. For simplicity, we use only the data collected by the S3 chip of the ACIS-S (Advanced CCD Imaging Spectrometer)  or by the I0, I1, I2, and I3 chips of the only ACIS-I observation. The data are reduced with the \chandra\ Interactive Analysis of Observations  \citep[CIAO v.4.13;][]{2006SPIE.6270E..1VF}, using $Chandra$ Calibration Database (CALDB v4.9.5). 
\begin{figure*}
\gridline{\fig{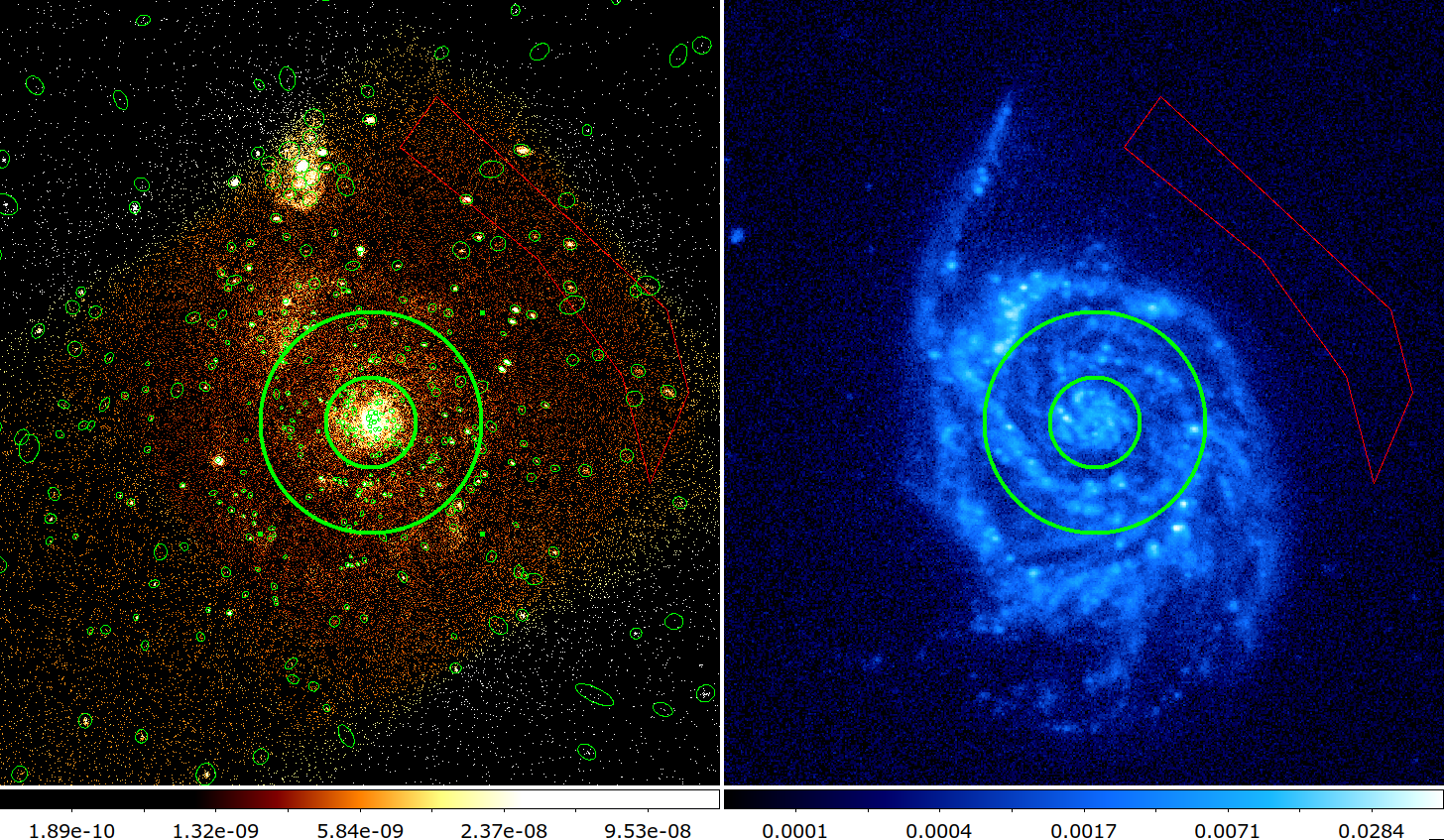}{1\textwidth}{}}
\caption{Overview of M51: The left panel shows the count intensity image of the combined $Chandra$ data in the 0.5-7.0~keV band, while the right panel shows the $GALEX$ FUV intensity image for comparison. The spectral analysis in the present work is focused on the X-ray data within the annulus outlined by the two large green circles with their radii equal to 45\as\ and 110\as\ in both panels.  Also shown is the red polygon enclosing the region where the local background of the galaxy is estimated. Those green ellipses in the left panel represent detected discrete source regions, which are masked out for the analysis of the diffuse X-ray emission. \label{f:observation}}
\end{figure*}

\begin{deluxetable}{ccccc}
\tablecaption{Chandra observations}
\tablewidth{0pt}
\tablehead{
\colhead{ObsID} & \colhead{Date}  & \colhead{Detector} &
 \colhead{Mode\tablenotemark{a}} & \colhead{Exp} \\
 &  & &
  & \colhead{(ks)} \\
}
\startdata
354 & 2000-06-20 & ACIS-S & F & 14.86\\
1622 & 2001-06-23 & ACIS-S & VF & 26.81 \\
3932 & 2003-08-07 & ACIS-S & VF & 47.97 \\
12526 & 2011-06-12 & ACIS-S & VF & 9.63 \\
12668 & 2011-07-03 & ACIS-S & VF & 9.99 \\
13812 & 2012-09-12 & ACIS-S & F & 157.46 \\
13813 & 2012-09-09 & ACIS-S & F & 179.2 \\
13814 & 2012-09-20 & ACIS-S & F & 189.85 \\
13815 & 2012-09-23 & ACIS-S & F & 67.18 \\
13816 & 2012-09-26 & ACIS-S & F & 73.10 \\
15496 & 2012-09-19 & ACIS-S & F & 40.97 \\
15553 & 2012-10-10 & ACIS-S & F & 37.57 \\
19522 & 2017-03-07 & ACIS-I & F & 37.76 \\
20998 & 2018-08-31 & ACIS-S & VF & 19.82 \\
23472 & 2020-10-13 & ACIS-S & F & 33.62 \\
23473 & 2020-11-18 & ACIS-S & F & 34.51 \\
23474 & 2020-12-21 & ACIS-S & F & 36.14 \\
23475 & 2021-01-28 & ACIS-S & F & 34.51 \\
23476 & 2021-03-01 & ACIS-S & F & 34.44 \\
23477 & 2021-04-01 & ACIS-S & F & 31.64 \\
23478 & 2021-05-04 & ACIS-S & F & 31.55 \\
23479 & 2021-06-07 & ACIS-S & F & 35.00 \\
23480 & 2021-07-31 & ACIS-S & F & 34.51 \\
23481 & 2021-08-18 & ACIS-S & F & 17.72 \\
25689 & 2021-08-19 & ACIS-S & F & 16.90 \\
\hline
total &            &         &  &  1252.71 	 \\
\enddata
\tablenotetext{a}{F = ``Faint'', VF = ``Very Faint''}
\label{t:observations}
\end{deluxetable}

We start by reprocessing the level 1 (raw) event data to generate a new level 2 event file for each observation, using $chandra\_repro$. We merge the data to create the combined count,  exposure, and flux maps in the soft (0.5-2 keV), hard (2-7 keV), and broad (0.5-7 keV) bands. For discrete source detection and removal in each band, we produce a 90\% energy-encircled region (EER) map of the point spread function (PSF) by the exposure-weighted mean of the corresponding images of individual observations. 

\subsection{Discrete X-ray sources: detection, removal, and residual contribution} \label{sec:point sources removal}

Since this work focuses on diffuse emission, our source detection is more inclined towards achieving completeness rather than reducing false detections. {\small WAVEETECT} is used to detect discrete sources on the scales of 1, 2, 4, and 8 pixel sizes and with the false detection probability of $< 10^{-7}$ in the three bands. 
Source detection is performed first on individual observations. The source-removed event data and the routine {\small LC$\_$CLEAN} are then used to remove time intervals with significant background flares.  
We correct for offsets of individual observations relative to the longest observation (ObsID: 13814), using {\small REPROJECT$\_$ASPECT}. After this alignment and data merging, we rerun the source detection, repeat it after removing events in each source region, and exclude any newly detected sources, which, if located near a bright source, could be missed in the initial detection. Detections in different bands are considered to be the same source if their 90\% EER regions overlap. The size and position of an adopted source region are determined by the {\small WAVEETECT} detection, selected according to the band priority: broad> soft> hard. We adopt a removal region of each source as an ellipse whose major and minor axes are linearly 3 times the 90\% EER.  
For visualization only, we fill the holes left from source removal, using {\small DMFILTH}, which interpolates the event density estimated in the immediate vicinity.

We include the residual contribution from discrete X-ray sources in our spectral modeling. Part of this contribution arises from the incomplete removal of the detected sources, due to their PSF wings beyond the removal regions and to the sources below our detection limit ($\approx 1.6 \times 10^{36} {\rm~erg~s^{-1}}$in the 0.5-7.0~keV band at the M51 distance).  In the active star-forming galaxy M51, they should typically be high-mass X-ray binaries with a rather flat luminosity function. The emission should be dominated by such binaries just below the limit and is unevenly distributed spatially. Therefore, it is better to model the contribution locally in individual regions of the galaxy. To do so, we characterize the contribution with a power law with the photon index fixed at 1.4 \citep{Lumb2002}. 

In addition, we expect a small collective contribution of unresolved X-ray emission from the old stellar population of M51. Dominated by low-mass X-ray binaries (LMXBs), cataclysmic variables (CVs), and coronally active binaries (AB), this contribution can be estimated from the stellar mass of the galaxy. \citet{Lin2015} estimated the X-ray luminosity function of LMXBs from long exposure of NGC 3115, where the normalization factor is proportional to the stellar mass. \citet{Ge2015} built the relationship between stellar mass and X-ray luminosity of CVs and ABs from Local Group dwarf elliptical galaxies. Following their work, we estimate the stellar mass, using the 2MASS K band data, and then the 0.5-2.0~keV luminosity of this contribution should be lower than $7 \times 10^{37} {\rm~erg~s^{-1}}$ in the annulus considered here (Fig.~\ref{f:observation}; see \S~\ref{sec:spiral gridding}). The luminosity is about two orders of magnitude lower than that observed in the same region due to our low detection limit(\S~\ref{sec:result}). For simplicity, we do not explicitly include the contribution in our spectral modeling.

\begin{deluxetable}{lr}
\tablewidth{0pt}
\tablenum{3}
\tablecaption{Spectral analysis results of the X-ray background}
\tablehead{
\colhead{Parameter} & \colhead{Value}  
}
\startdata
kT\(_1\) (keV) & 0.1 keV (fixed) \\
Z\(_1\) (solar) & 1 (fixed) \\
EM\(_{l,1}\) (10\(^{18}\)  cm\(^{-5}\)) & \(< 0.01\) \\
N\(_{H}\) (10\(^{20}\)  cm\(^{-2}\)) & 3.3 (fixed) \\ 
kT\(_2\) (keV) & \(0.18^{+0.04}_{-0.01}\) \\
Z\(_2\) (solar) & \(< 0.1\) \\
EM\(_{V,2}\) (10\(^{18}\) cm\(^{-5}\)) & \(1.08^{+0.03}_{-0.65}\) \\
\(\Gamma_{po}\) & 1.4 (fixed) \\
norm/A\(_{po}\)\tablenotemark{a} & \(6.2^{+1.2}_{-1.3}\) \\
\(\chi^2 / \text{d.o.f.}\) & 52/67 \\
\enddata
\tablenotetext{a}{Normalization per area in unit of \(10^{-51}\) keV\(^{-1}\) cm\(^{-4}\) s\(^{-1}\) at 1 keV.}
\label{t:background}
\end{deluxetable}

\subsection{Background estimation and subtraction} \label{sec:backgrund subtraction}
The background consists of components of non-cosmic and cosmic events. The non-cosmic component is estimated by the software {\small mkacispback}, introduced by \citet{Suzuki2021} and based on a model constructed using data from the ACIS-stowed and Chandra Deep Field South survey. This estimation is first performed for individual observations and then combined with effective exposures as weights. 

We estimate the X-ray background locally to minimize the non-negligible cosmic variance, e.g., due to the Galactic foreground that is especially important at $\lesssim 1$~keV, following the strategy described in \citet{Cheng2021}. Briefly, we extract a local background spectrum from the background region shown in Fig.~\ref{f:observation} and fit it (after subtracting the non-cosmic component) with the XSPEC model APEC+TBABS*(APEC+POWERLAW) \citep[see Appendix A of][]{Cheng2021}. The best-fit model as shown in Table \ref{t:background} is used to predict the spectral component of the local cosmic X-ray background across the field of M51, as well as the count intensities in the individual bands for imaging analysis. However, one should not over-interpret the numbers in Table \ref{t:background}. The model is designed only to characterize the overall shape of the background spectrum. The fitted parameters are degenerate with each other and are not necessarily physical. This makes comparing the parameters themselves fitted in different studies not particularly meaningful. 
We combine the two components to estimate the total background contributions in individual regions. 
\begin{figure*}
\gridline{\fig{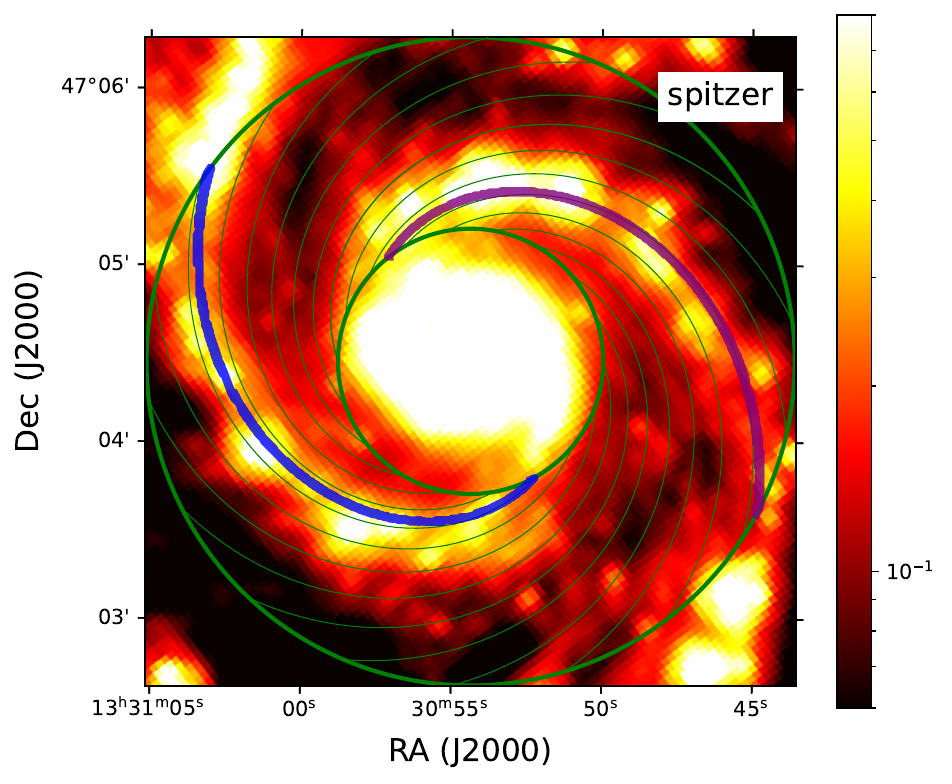}{0.5\textwidth}{}
\fig{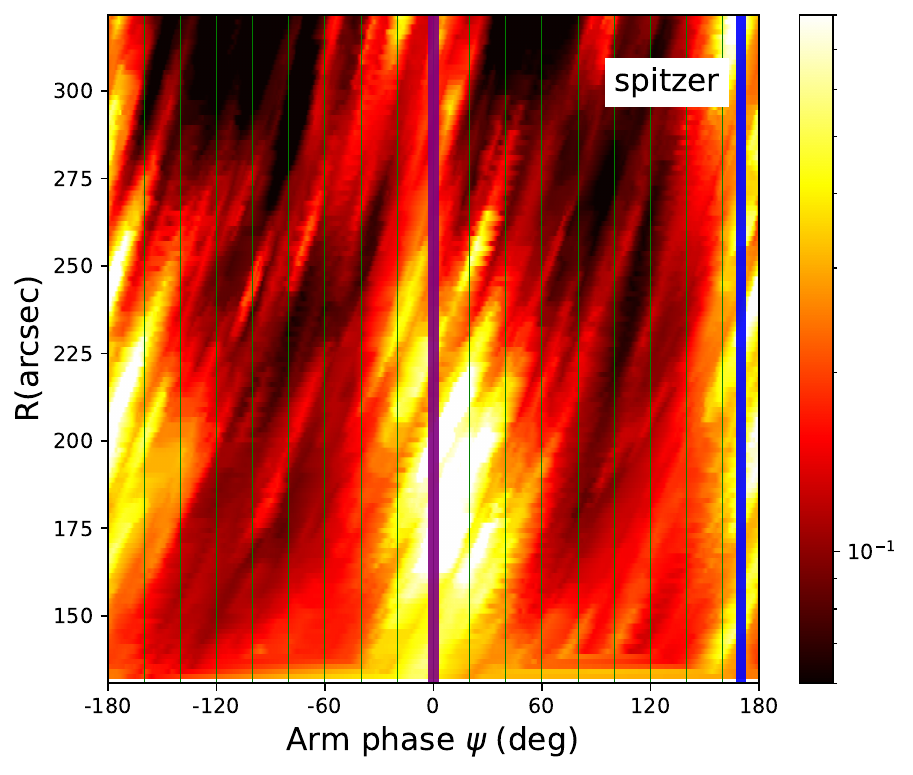}{0.5\textwidth}{}}
\gridline{\fig{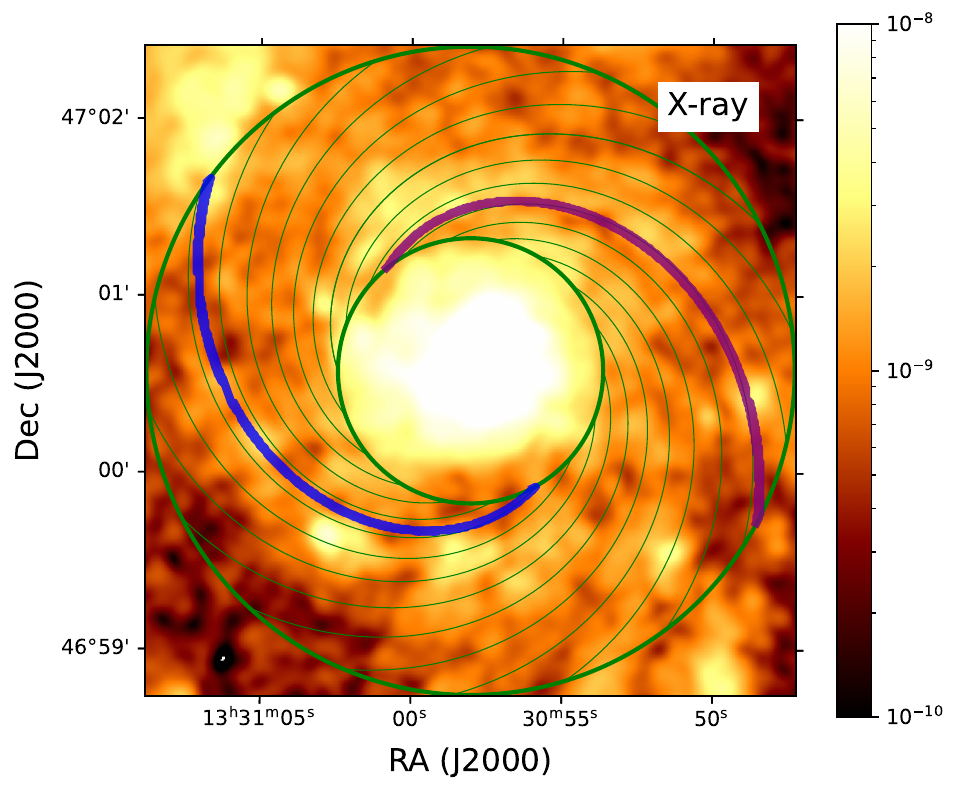}{0.5\textwidth}{}
\fig{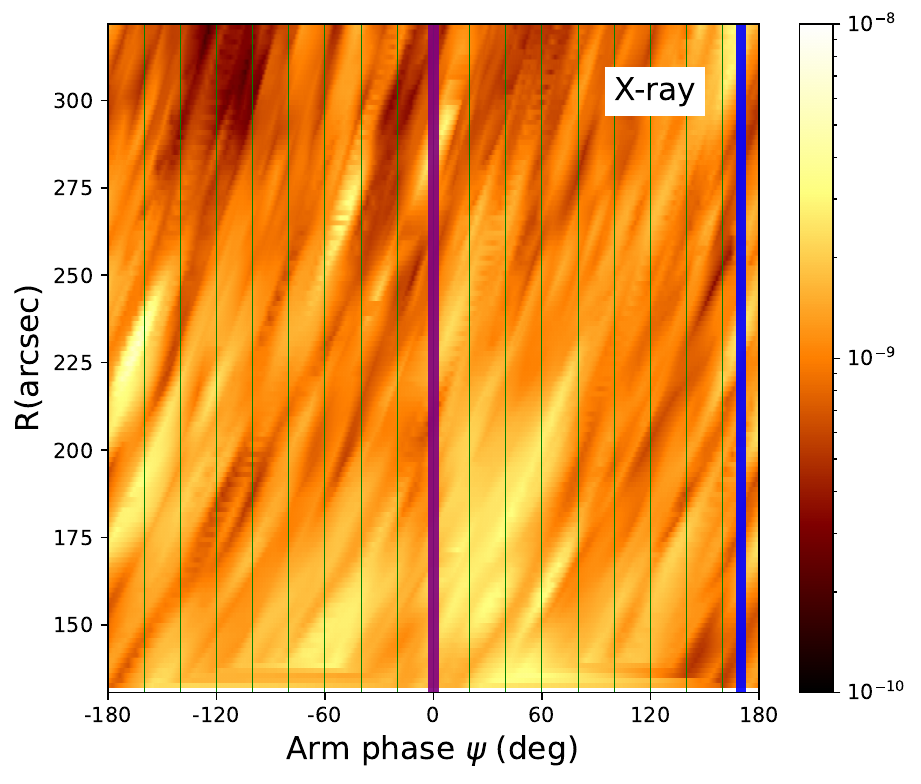}{0.5\textwidth}{}}
\caption{Viewing M51 in spiral arm phases: {\sl Spitzer} 24 \micron\ intensity  (top panels) and smoothed $Chandra$ ACIS-S 0.5-2.0~keV intensity (bottom panels) in the Cartesian (left) and logarithmic curvilinear (right; see Eq.~\ref{e:coor}) coordinates. The purple and blue solid lines generally represent the front position of the two spiral arms in all panels. The arm phase is relative to the purple arm and is positive/negative in the anti-clockwise/clockwise direction. The two green circles in the respective left panels are the same as those in Fig.~\ref{f:observation}, outlining the annulus, which is divided into 18 regions by thin green logarithmic spiral curves. We extract X-ray spectra from these 18 regions (see Section \ref{sec:extraction}). The right panels, generated from the data in the annulus via interpolation, are only for visualization. }\label{f:gridding}
\end{figure*}

\subsection{Spiral curvilinear coordinates}\label{sec:spiral gridding}
The dominant structure of M51 is, of course, the two grand design spirals. Following \citet{Shetty2007}, \citet{Koda2012} and \citet{Pineda2020}, we define a 2-D curvilinear coordinate system, in which one axis (or arm phase $\psi$) is always perpendicular to a spiral arm and the other (the radius $R$) is parallel to it. We also add a correction for the disk inclination angle. If the coordinates of a point are ($r,\theta$) in the polar coordinate system with the galaxy center as the pole and north as the pole axis, then after the coordinate system transformation, its coordinates in the spiral arm coordinate system are ($R,\psi$) as
\begin{subequations}
\begin{equation}
            R = \frac{r}{{\rm sin}(i_{pitch})} \sqrt{1-{\rm sin}^2(\theta-\theta_{PA}){\rm tan}^2(i_{inc})}
\end{equation}
\begin{equation}
\psi = {\rm arctan}[{\rm tan}(\theta-\theta_{PA}){\rm sec}(i_{inc})] + \frac{1}{{\rm tan}(i_{pitch})} {\rm ln}(\frac{r}{r_{min}}),
\end{equation}\label{e:coor}
\end{subequations}
where $i_{pitch}$ is the pitch angle of the spirals, $\theta_{PA}$ is the position angle, $i_{inc}$ is the inclination angle, $r_{min}=45 ^{\prime\prime} $, which appears in the formula to ensure that the spiral curve with $\psi=0^\circ$ passes through the point ($\theta=\theta_{PA}, r=r_{min}$). 
Both arms in M51 roughly follow the logarithmic form with  $i_{pictch}=18.5^\circ$ at  radii  $r \lesssim 2'$ . However, the pitch angle decreases to $i_{pictch}=2^\circ$ at radii range $2' \lesssim r \lesssim 3'$and then rapidly increases to approximately $28^\circ$ starting from $r \gtrsim 3'$ \citep{Pineda2020}. This change in pitch angle indicates that the spirals are probably not stable, as may be caused by the interaction of M51 with NGC\,5195 (M51b) \citep{Pineda2020}, which is expected to have the most significant impacts at relatively large radii. 
For simplicity, we focus on the region within the annulus from radius 45\as\ to 110\as, where the data quality of the diffuse X-ray emission is also the highest. Table \ref{t:M51} lists the key parameters used in the coordinate system transformation and other calculations, while Fig.~\ref{f:gridding} demonstrates the 2-D transformation result.

\subsection{Spectral extraction}\label{sec:extraction}
The decent counting statistics of the M51 allow us to divide the annuls into 18 different spiral arm-phase regions, each of which spans $\delta \psi = 20^\circ$. For each region, we first extract spectra from individual observations using $specextract$ and then combine them using $combine\_spectra$. The channels of the combined spectrum are adaptively grouped to ensure that each bin has a signal-to-noise ratio $S/N \large >3$, where the signal is the number of source counts after background subtraction. The spectra of the regions are jointly analyzed to examine the spiral-arm phase dependence of the X-ray emission and absorption properties.  In addition, we produce an integrated spectrum of the annulus by combining the 18 regional spectra. This integrated spectrum is only used to concisely illustrate the average spectral shape, compared to the best-fit model to the individual regional spectra (see Section \ref{sec:result}).

\subsection{Hierarchical Bayesian approach of the spatially-resolved spectral analysis}

Our goal is to meaningfully constrain the spectral properties of the diffuse hot plasma, not only in the $N=18$ individual regions but also over the entire annulus. 
A commonly used approach is to assume certain (global) model parameters to be the same for all the regions while estimating other (local) parameters by fitting them to individual spectra. This approach may be realized easily with a standard (non-hierarchical) Bayesian analysis, following Bayes' theorem: 
\begin{equation}
\begin{split}
p (  \mathbf{\alpha}, \mathbf{\gamma}_i \vert \mathbf{I}_\epsilon  ) \propto p(\mathbf{\alpha} )*\displaystyle\prod_{i } p ( \textbf{I}_\epsilon \vert \mathbf{\alpha}, \mathbf{\gamma}_i )  *p(\gamma_i),\\
\end{split}
\end{equation}
where $\alpha$ represents the vector composed of the global parameters, $\gamma_i$ is the vector composed of the local parameters to be fitted to individual spectra $i$ (from 1 to $N$),  and ${\bf I_\epsilon}$ is the spectral data set, while the probability distribution priors, $p(\alpha)$ and $p(\gamma_i)$, are usually set to be uniform over physically plausible ranges. However, this simple approach is often not ideal. For a local parameter, its constraint only comes from one spectrum, which usually has too limited counting statistics. For a global parameter,  we learn nothing about its intrinsic distribution or any variation trend from one region to another. 

Here, we instead adopt the so-called hierarchical Bayesian analysis approach. It allows us to characterize the similarity or intrinsic distribution of local model parameters and/or take advantage of existing knowledge or physical intuition in the spectral modeling. Specifically, we prescribe a distribution function as a hyper-prior for an interesting local parameter. This function may contain so-called hyperparameters that can be fitted in the analysis, following the modified Bayes theorem:

\begin{equation}
\begin{split}
 &p (  \mathbf{\alpha},\mathbf{\beta},\mathbf{\gamma}_i, \mathbf{\omega}_i\vert \mathbf{I}_\epsilon  )  \\
 &\propto  p(\mathbf{\alpha})* \displaystyle\prod_{i }p ( \textbf{I}_\epsilon \vert \mathbf{\alpha}, \mathbf{\gamma}_i , \mathbf{\omega}_i )*p(\mathbf{\gamma}_i)*p(\mathbf{\omega}_i)\\
 &\propto  p(\mathbf{\alpha})* p(\mathbf{\beta})*\displaystyle\prod_{i }p ( \textbf{I}_\epsilon \vert \mathbf{\alpha}, \mathbf{\gamma}_i , \mathbf{\omega}_i )*p(\mathbf{\gamma}_i)*p(\mathbf{\omega}_i\vert \mathbf{\beta})\\
\end{split}
\end{equation}
where the new vector $\mathbf{\omega}_i$ represents the local parameters constrained by the hyperpriors determined by the hyperparameter vector $\mathbf{\beta}$. Not all local parameters need to have a hyperprior, which can be computationally intensive and requires a good justification. We designate those completely free local parameters as the vector $\mathbf{\gamma}_i$ and the global parameters that are the same for all regions as $\mathbf{\alpha}$. 

The use of the hyperprior is intended to provide constraints on a local parameter that is statistically similar from one region to another.  In principle, the form of the prior and its hyperparameters may be determined independently, e.g., via empirical measurements and computer simulations. Absent of such knowledge for a local parameter, one may just intuitively assume a form for its hyperprior with the hyperparameters fitted in the analysis.  The suitability of this assumed form can be statistically tested by comparing it with the parameter's posterior distribution averaged over the fitted regions. Specifically, we assume that the hyperprior follows a normal distribution,
\begin{equation}
\begin{split}
&p(x_i\vert \mu_{x},\sigma_{x})=\frac{1}{\sqrt{2 \pi} \sigma_{x}}{\rm exp}\left[-\frac{(x_i-\mu_{x})^2}{2\sigma_{x}^2}\right],\\
\end{split}\label{e:kt-prior}
\end{equation}
where the mean $\mu_x$ and the variance $\sigma_x$ are two hyperparameters to be fitted, while the parameter  $x_i \equiv {\rm log}(kT_i)$ is the local parameter for a region $i$ in Models.

\begin{figure*}
\gridline{\fig{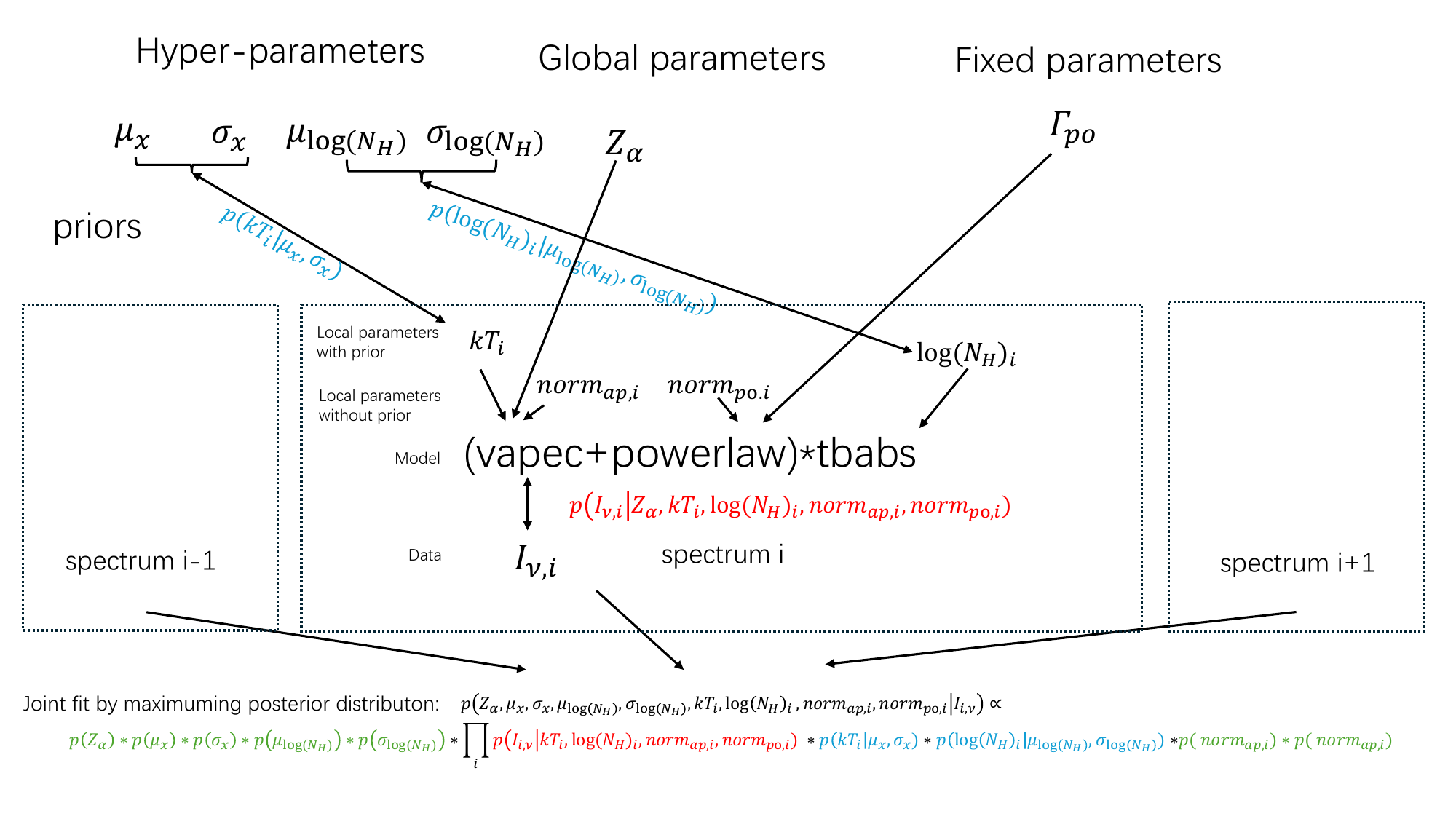}{1\textwidth}{}}
\caption{Illustration of the hierarchical structure of our spectral modeling, taking Model A (see Table~\ref{t:spec}) as an example. The part of the structure inside each dashed box (although only Spectrum i is detailed) includes local parameters and modeling steps (generally defined in the caption to Table~\ref{t:spec}) and executed in Xspec. The part outside the box consists of global parameters and hyperpriors (all assumed to have a normal distribution form), with their parameters to be fitted jointly. Finally, the posterior probability consists of three parts: the likelihood (red), the priors on the local parameters with the hierarchical structure (blue), and the priors on the local parameters without the hierarchical structure (green, which are usually assumed to be flat). Maximizing the posterior probability via the MCMC random walk leads to constraints on all the fitting parameters.  
\label{f:ill}}
\end{figure*}

We implement the hierarchical Bayesian analysis approach on top of XSPEC (\citet{1996ASPC..101...17A}; but v12.12 ) -- a spectral analysis software package. Although allowing only the standard Bayesian approach, the Python version of XSPEC, or pyxspec, can return the goodness-of-fit probability at each step of the analysis.
We use this probability, together with our MCMC walking algorithm and implementation of the hyperpriors, to complete the analysis. The results of the analysis for various spectral models (Fig.~\ref{f:geo_diagram}) are detailed below in \S~\ref{sec:result}, while Fig.~\ref{f:ill} gives an example of the hierarchical structure of such an analysis.

\subsection{Plasma model with a lognormal temperature distribution}\label{s:lognormal_model}
Similarly to the finding in \citet{Cheng2021} and \citet{Zeng2023} for the starburst region 30 Doradus and the face-on spiral galaxy M83, we find that the simple 1-T optically thin thermal plasma model $vapec$ cannot give a satisfactory fit to some individual spectra of good counting statistics, indicating that the temperature distribution of diffuse hot plasma needs to be taken into account. Thus, we test the spectral model $vlntd$, assuming a lognormal temperature distribution. The implementation of this model follows that described by \citet{Cheng2021} in their appendix D. However, we change the variable $x$  from $x \equiv {\rm ln}(kT)$ to $x \equiv {\rm log}(kT)$ or a logarithmic variable based on 10.  This change makes it easier to directly use and understand the fitted parameters (now simply in dex) for a region $i$, $x_i \equiv {\rm log}(kT_i)$ and $s_{i}$, or the mean and dispersion of $x$, where $kT_i$ is the median temperature (in units of keV). For example,  $x_i=-1$ and $s_i=1$ would suggest that the plasma in this region has a median temperature of 0.1\,keV and a 1 $\sigma$  uncertainty range from 0.01\,keV to 1.0\,keV. More specifically,  the model is the sum of the spectral contribution from multi-temperature plasma weighted by the emission measure (EM) distribution, 
\begin{equation}
\begin{split}
\frac{dEM}{dx}=&\frac{1}{\sqrt{2 \pi} s_{i}}{\rm exp}\left[-\frac{(x-x_i)^2}{2s_{i}^2}\right].\\
\end{split}\label{e:kt-local}
\end{equation}

We emphasize that Eq.~\ref{e:kt-prior} is a prior describing the relationship between the hyperparameters ($\mu_x$ and $\sigma_x$) and a set of local parameters ($x_i$) that are used to characterize the spectra of individual regions $i$. The hyperparameters are determined by the model fitting to the spectra of these multiple regions. Thus, $\mu_x$ is the global mean of $x$ (e.g., across much of the galaxy), while $\sigma_x$ characterizes the variation between the regions. In contrast, Eq.~\ref{e:kt-local} is only part of the model to fit the spectrum of {\sl one} region; the parameters $x_i$ and $s_i$ describe the internal or small-scale temperature distribution that is not spatially resolved in the spectral modeling.

\subsection{Multiplicative model of the X-ray absorption mixed with the emission}

We implement this model as $mxabs$ for use in Xspec. The model assumes that the X-ray absorbing medium is mixed with the X-ray emission macroscopically in a uniform fashion, that is, the ratio of volume emissivity to absorption coefficient, or $j_\epsilon/\kappa_\epsilon$  (at energy $\epsilon$), is constant. For simplicity, we have neglected scattering or self-absorption of the X-ray-emitting plasma. The corresponding radiation transfer equation can be expressed as
\begin{equation}
\begin{split}
\frac{d I_\epsilon(s)}{ds} = j_\epsilon (s)-\kappa_{\epsilon}(s) I_\epsilon(s),\\
\end{split}
\end{equation}
where $ I_\epsilon$ is the specific X-ray intensity along the path $s$.  Assuming no background radiation, this equation can be solved as 
\begin{equation}
\begin{split}
I_\epsilon & = \frac{j_\epsilon}{\kappa_{\epsilon}} (1-e^{-\tau_{\epsilon}}),\\
& = I_{\epsilon,0} \frac{ 1-e^{-\tau_{\epsilon}} }{\tau_{\epsilon}}, \\
\end{split}
\end{equation}
where $\tau_\epsilon= \int \kappa_{\epsilon} ds$ is the total opacity integrated along the line of sight through the emission region, while $I_{\epsilon,0} = \int j_\epsilon ds = \frac{j_\epsilon}{\kappa_{\epsilon}} \tau_\epsilon$ is the total X-ray intensity without accounting for the absorption. We find that the model is the same as the intrinsic multilayer absorption model described by \citet{Fabian2022}, although their derivation is somewhat different.

\subsection{Neglecting charge exchange contribution}\label{ss:CX}

In recent years, charge exchange (CX) between hot plasma and cool gas has been shown to contribute significantly to the soft X-ray emission in nearby star-forming galaxies \citep[e.g.,][]{Liu2011,Liu2012,Zhang2014,Lopez2020}.  This contribution has been demonstrated in the analysis of an \xmm/Reflection Grating Spectrometer (RGS) spectrum of M51 \citep{Zhang2022}.  This spectrum, covering the bright diffuse X-ray emitting region at the northeastern corner of Fig.~\ref{f:geo_diagram} left panel, shows a high G ratio of the OVII He$\alpha$ triplet lines [$=(f+i)/ r \approx 1.4$, where the resonance ($r$) line is at 574 eV, the intercombination ($i$) line is at 569 eV, and the forbidden ($f$) line is at 561 eV], which is not expected for the thermal plasma in a collisional ionization equilibrium and at temperatures $\gtrsim 0.1$~keV, but can naturally be explained by the CX. Its contribution is qualitatively significant, although its exact fraction depends on the spectral modeling. The region covered by the RGS data is just outside the annulus considered in the present work. Nevertheless,  \citet{Zhang2022} show that the CX contribution is consistent with the \chandra\ ACIS-S data across the galaxy, based on the 1-T plasma modeling and assuming the foreground-only absorption.  Including the CX contribution in a more sophisticated modeling of the plasma and absorption, as is focused in the present work, remains difficult. The contribution is important only at the very soft end of the observed spectra ($\lesssim 0.6$~ keV) and has a limited effect on the measurements of the spectral parameters. Therefore, for simplicity and for an independent test of its importance, we largely neglect the contribution of the CX in the spectral modeling (but see the discussion at the end of \S~\ref{sec:result}). 
\section{Results}\label{sec:result}

\begin{figure*}
\gridline{\fig{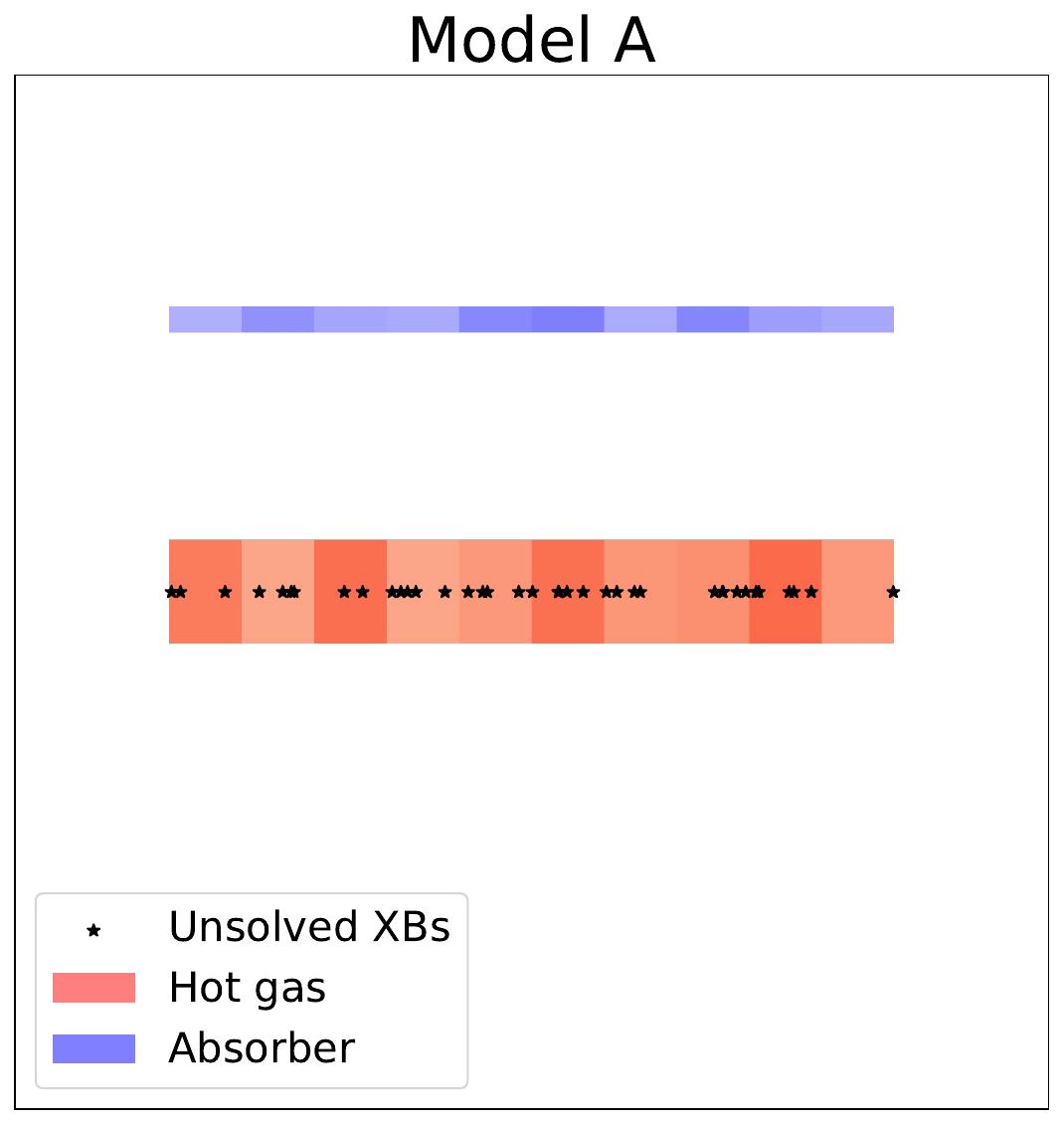}{0.25\textwidth}{} 
\fig{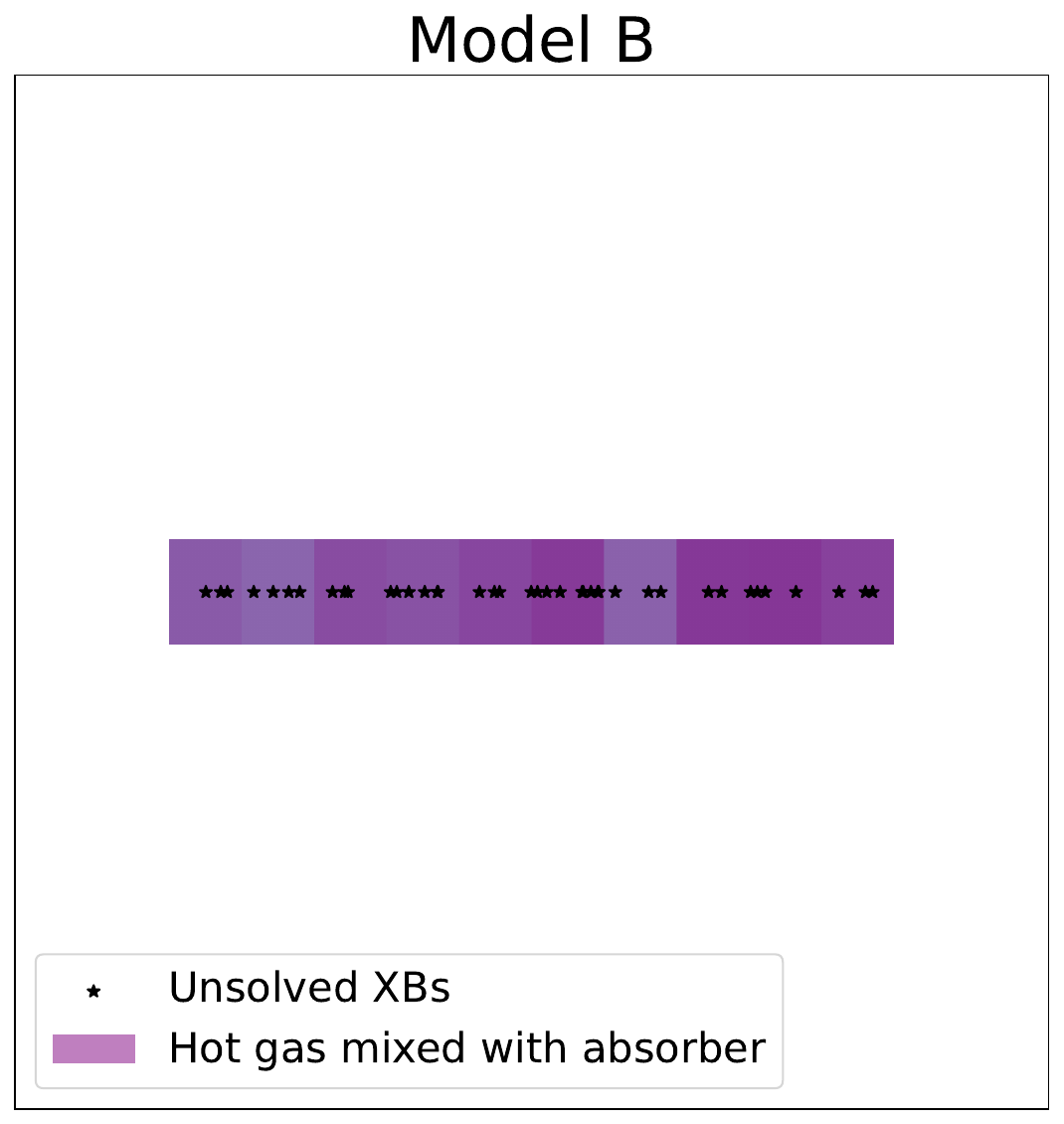}{0.25\textwidth}{} 
\fig{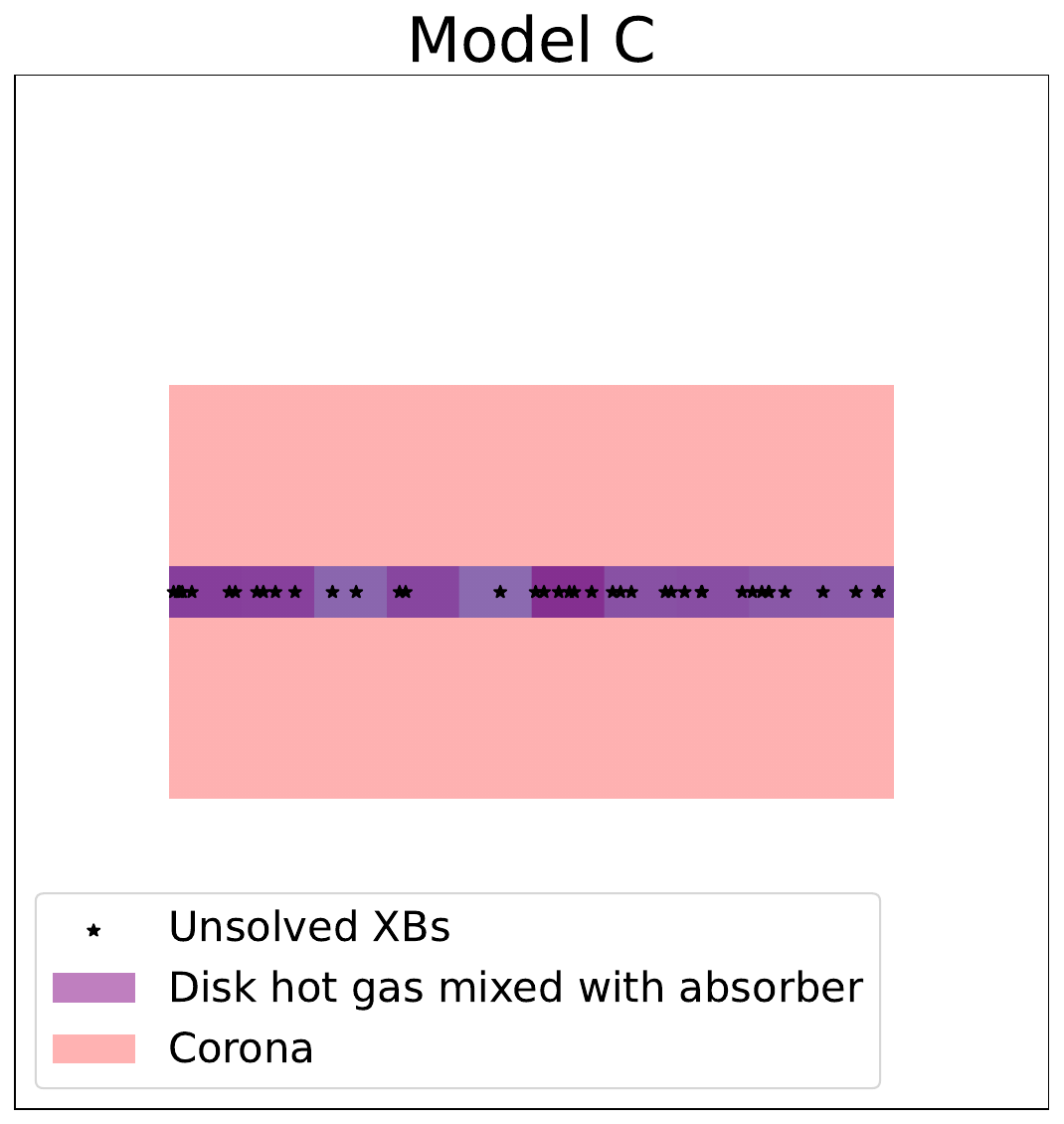}{0.25\textwidth}{} 
\fig{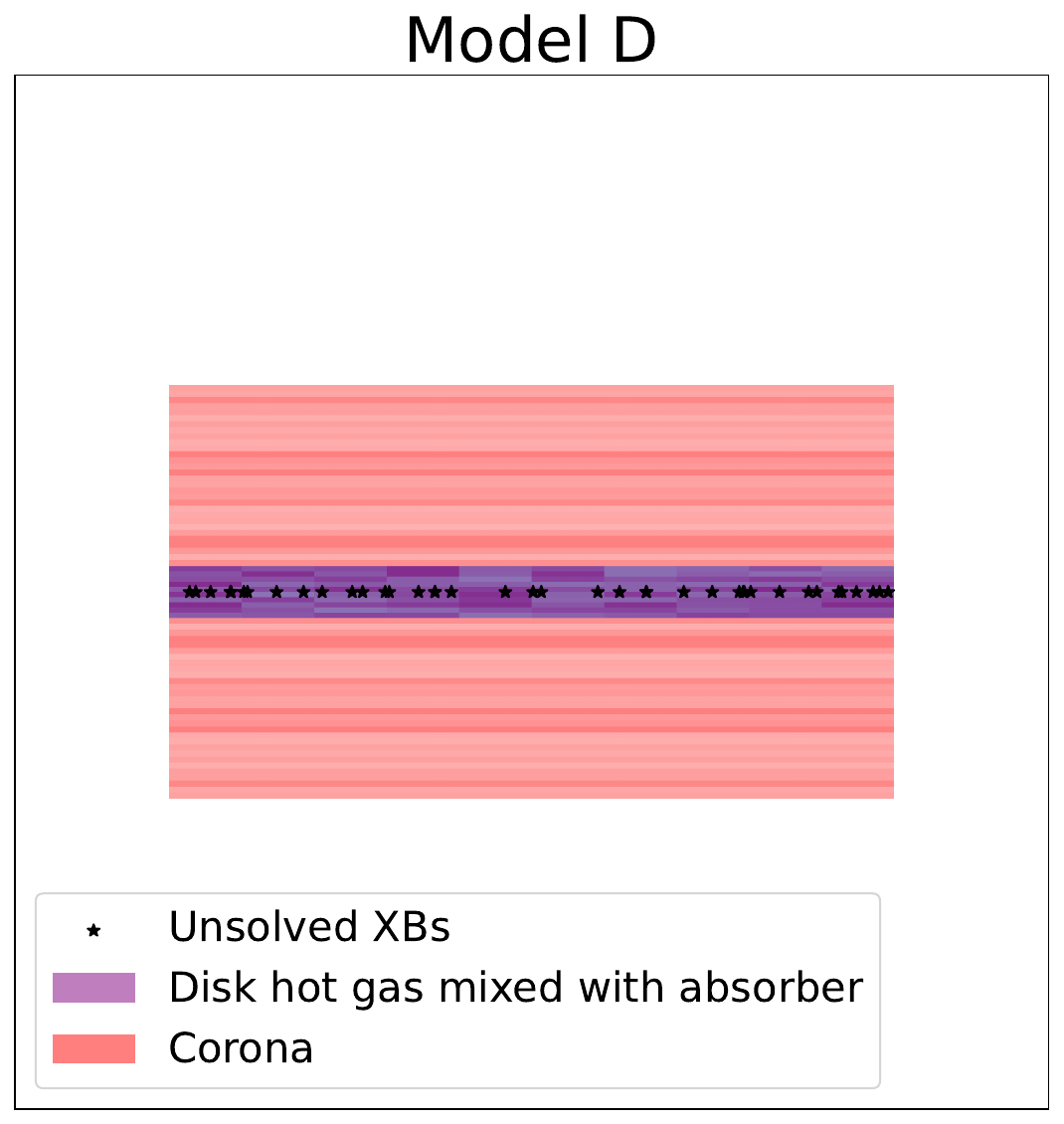}{0.25\textwidth}{} 
}
\caption{Illustrative geometric constructions of the four spectral models presented in this paper.  
Here we have chosen an edge-on perspective, although M51 is a face-on galaxy. The horizontal color change illustrates the variation of the hot plasma and/or absorbing gas properties with the spiral arm phase. Variation of the color for the hot plasma in Model D indicates the vertical non-uniformity of  its temperature. The putative observing direction is from the top.} \label{f:geo_diagram}
\end{figure*}

Fig.~\ref{f:geo_diagram} shows the geometry of the four models presented here. Table~\ref{t:spec} lists the best-fit values and 1$\sigma$ uncertainties for the hyperparameters and global parameters of these models.
Each model also includes local parameters, which are constrained primarily by individual spectra. Such constraints are relatively weak and are typically less quantitatively interesting.  However,
the fitted parameter values can collectively show potential trends among the regions or correlation with other stellar and ISM properties.  All models are subject to Galactic foreground absorption ($tbabs_G$) with the column density as given in Table~\ref{t:M51}. Considering the limited spectral resolution of the \chandra\ ACIS-S data, we use the single parameter  $Z_\alpha$, or the abundance ratio of $\alpha$  to Fe-like elements, to characterize the abundance dependence of the hot plasma, while assuming the Fe abundance to be solar.

In the following, we describe the individual models and their fitting results.

\begin{deluxetable*}{lllcccc}
\tablenum{4}
\tablecaption{Results on the global model parameters of the spectral analysis}
\label{t:spec}
\tablewidth{0pt}
\tablehead{
\colhead{Model\tablenotemark{a}}&\colhead{} &\colhead{} & \colhead{A}  & \colhead{B}  & \colhead{C}  & \colhead{D}  \\
\colhead{Name} & \colhead{}& \colhead{}&\colhead{ Disk}  & \colhead{ Disk/\(N_H\)mixed}  &
 \colhead{ Corona+Disk/\(N_H\)mixed}  & \colhead{ log-T Corona+Disk/\(N_H\)mixed} }
\startdata
Component & Parameter & Note & & & & \\
\hline
 & $\mu_{x_d}$ & $b$ & $-0.835^{+0.006}_{-0.006}$ & $-0.529^{+0.007}_{-0.007}$ & $-0.10^{+0.11}_{-0.26}$ & $-1.00^{+0.11}_{-0.26}$ \\
 & $\sigma_{x_d}$ & $b$ & $0.005^{+0.001}_{-0.001}$ & $0.019^{+0.005}_{-0.004}$ & $0.05^{+0.02}_{-0.01}$ & -- \\
 Disk& $\mu_{s_d}$ & $c$ & -- & -- & -- & $0.53^{+0.09}_{-0.09}$ \\
 & $\sigma_{s_d}$ & $c$ & -- & -- & -- & $0.10^{+0.07}_{-0.04}$ \\
 & $Z_{\alpha,d}$ & $d$ & $0.049^{+0.006}_{-0.006}$ & $1.15^{+0.04}_{-0.04}$ & $1.79^{+0.22}_{-0.22}$ & $0.97^{+0.34}_{-0.32}$ \\
 \hline
 & $x_c$ & $b$ & -- & -- & $-0.747^{+0.010}_{-0.007}$ & $-1.07^{+0.07}_{-0.08}$ \\
Corona & $s_c$ & $c$ & -- & -- & -- & $0.54^{+0.04}_{-0.04}$ \\
 & $Z_{\alpha,c}$ & $d$ & -- & -- & $0.14^{+0.03}_{-0.03}$ & $1.40^{+0.14}_{-0.17}$ \\
 & $EM_{l,c}$ & $e$ & -- & -- & $1.17^{+0.19}_{-0.20}$ & $0.66^{+0.12}_{-0.09}$ \\
 \hline
Absorbing  & $\mu_{\rm log}(N_H)$ & $f$ & $21.69^{+0.02}_{-0.02}$ & $21.68^{+0.11}_{-0.18}$ & $21.15^{+0.11}_{-0.13}$ & $21.46^{+0.09}_{-0.09}$ \\
 gas (M51)& $\sigma_{\rm log}(N_H)$ & $f$ & $0.06^{+0.02}_{-0.01}$ & $0.34^{+0.07}_{-0.06}$ & $0.25^{+0.07}_{-0.05}$ & $0.11^{+0.05}_{-0.04}$ \\
 & $\chi^2$/d.o.f. & $g$ & 1261/1009 & 1430/1008 & 987/1006 & 941/1004 \\
\enddata 
\tablenotetext{a}{
Model definitions: \\
(A) \( (vapec+powerlaw)*tbabs_1\),\\
(B) \( vapec*mxabs_1+powerlaw*mxabs_2*tbabs_1\),\\
(C) \(vapec_1*tbabs_1+vapec_2+(vapec_3+powerlaw)*mxabs\), \\
(D) \( vlntd_1*tbabs_1+vlntd_2+(vlntd_3+powerlaw)*mxabs\).\\
Each model also includes a Galactic foreground absorption fixed at the value in Table~\ref{t:M51}. Only the fitted global and hyperparameters are listed here. Some local parameters of Model D are shown in Fig~\ref{f:TandnH2}.}
\tablenotetext{b}{ \(x={\rm log}(kT)\) is the 10-based logarithmic value of temperature in units  of \({\rm keV}\). \(\mu_x\)  and \(\sigma_x\) are the hyperparameters for \(x\) as defined in  Eq.~\ref{e:kt-prior}. Different from the other models, Model D assumes a lognormal temperature distribution. 
Since \(x_d\) and \(s_{d}\) are strongly degenerate in the fit to individual spectra,  we set \(x_d={\rm log}(kT_d)\) as a global parameter in this modeling (i.e., no \(\sigma_{x_d}\) is used).} 
\tablenotetext{c}{\(\mu_s\) and \(\sigma_s\) are the hyperparameters for \(s\) -- the dispersion of \(x\) (Eq.~\ref{e:kt-prior}). }
\tablenotetext{d}{ \(\rm{Z_{\alpha}}\) is the abundance of \(\alpha\) elements in solar pattern.
}
\tablenotetext{e}{ \(EM_l=\int n_e n_H dl\) (in the unit of \(10^{18}\ \rm{cm^{-5}}\)) is the column emission measure of the galactic corona on {\sl one side} of the galactic disk and is derived from \(norm\) of \(vapec\) or \(vlntd\). 
}
\tablenotetext{f}{ \({\rm log}(N_H)\) is the 10-based logarithmic value of H column density in the unit of \(\rm{cm^{-2}}\). \(\mu_{{\rm log}(N_H)}\) and \(\sigma_{{\rm log}(N_H)}\) are the hyperparameters for  \({\rm log}(N_H)\) as defined in Eq.~\ref{e:kt-prior}.
}
\tablenotetext{g}{In the hierarchical Bayesian analysis, the best fit value is obtained at the maximum posterior probability and does not exactly correspond to the  minimum \(\chi^2\) statistic. Nevertheless, the listed \(\chi^2\)/d.o.f. still provide a simple and useful goodness of fit check between the model and the data. The degree of freedom (d.o.f.),  conventionally defined as the number of data points minus the number of all fitting parameters, is also not strictly correct here,  because the local parameters are also constrained by the hyper-priors and are not completely independent in the spectral models for different regions. The numbers we provide here can be considered the most conservative estimate, or the minimum value of d.o.f., which is equal to the number of data points minus the number of all the global and local parameters.}
\end{deluxetable*}

\begin{figure}
\gridline{\fig{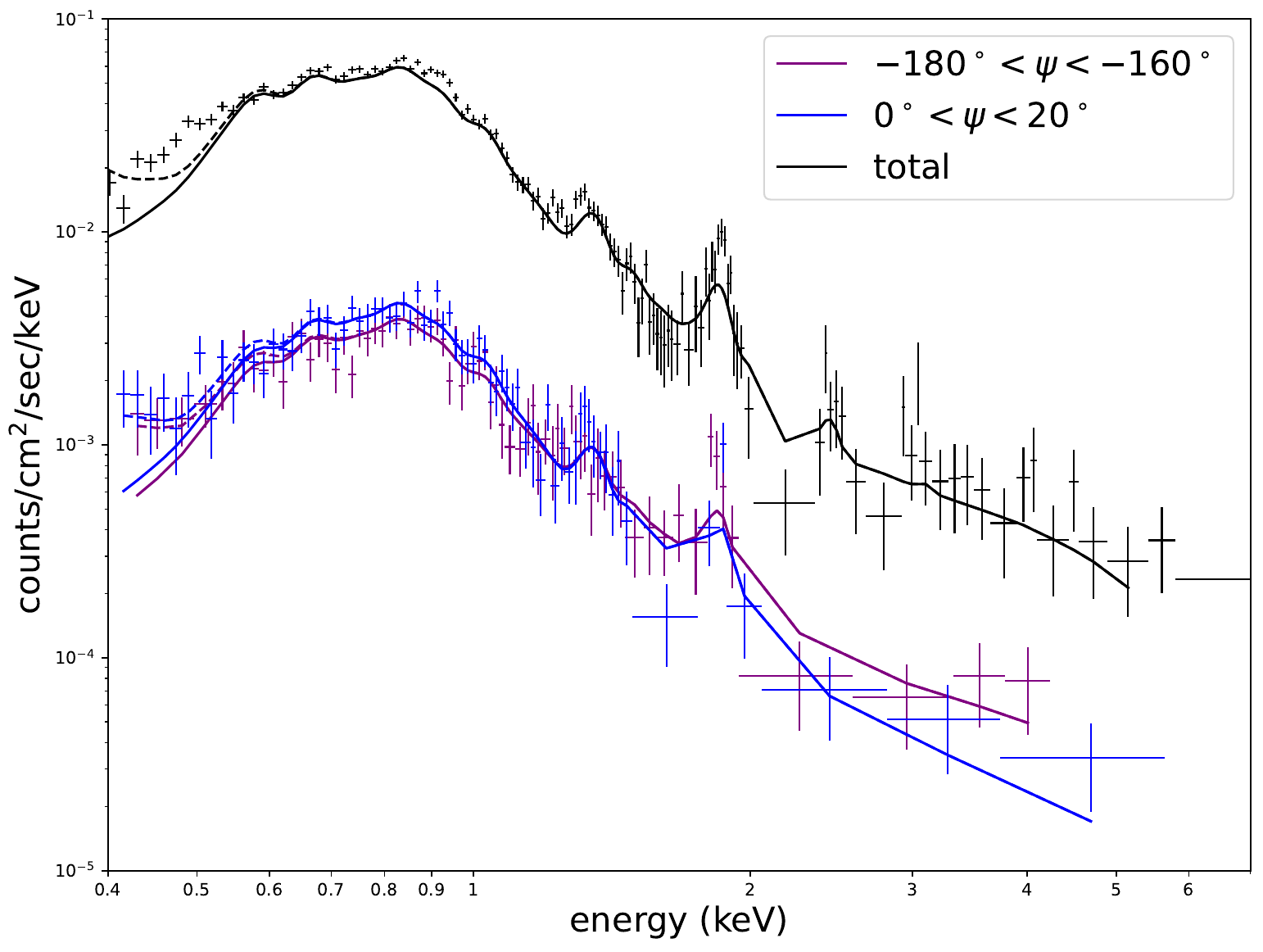}{0.5\textwidth}{}}
\caption{Sample \chandra\ ACIS-S spectra extracted from different arm phase regions, as well as the total combined spectrum of the annulus (Fig.~\ref{f:gridding}). The colors of the spectra are the same as the solid lines in Fig.~\ref{f:gridding}.  These spectra are all background-subtracted. The solid lines represent the result from fitting the spectra with Model D only (Table \ref{t:spec}), while the dashed lines are from the model fits including an approximate CX component (see Section \ref{ss:CX}.)}
\label{f:spectrum} 
\end{figure}

\begin{figure*}
\fig{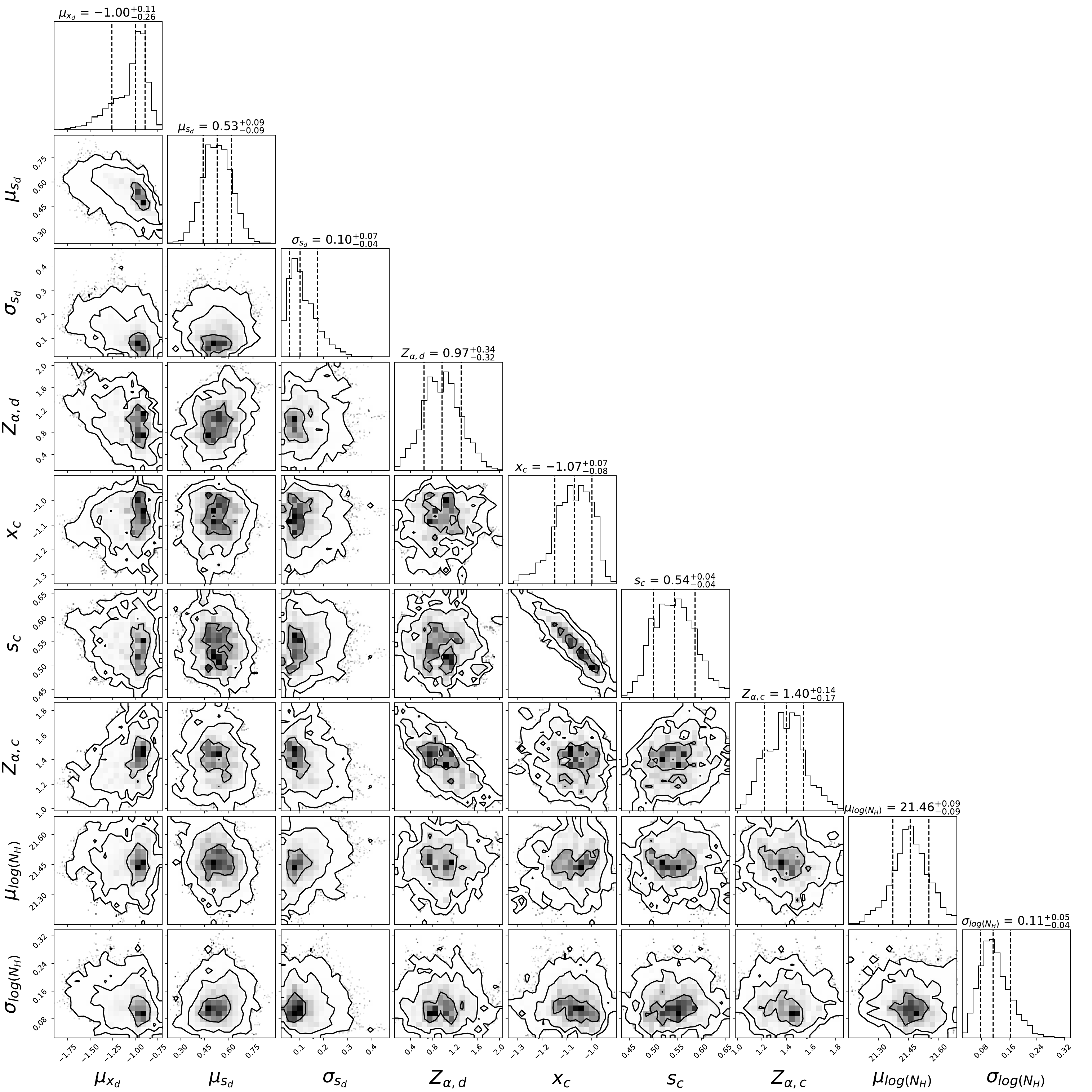}{1.0\textwidth}{}
\caption{Corner plot for the spectral fit with Model D: Posterior distributions of the global and hyper parameters.}\label{f:MCMC2}
\end{figure*}

\begin{figure*}
\gridline{\fig{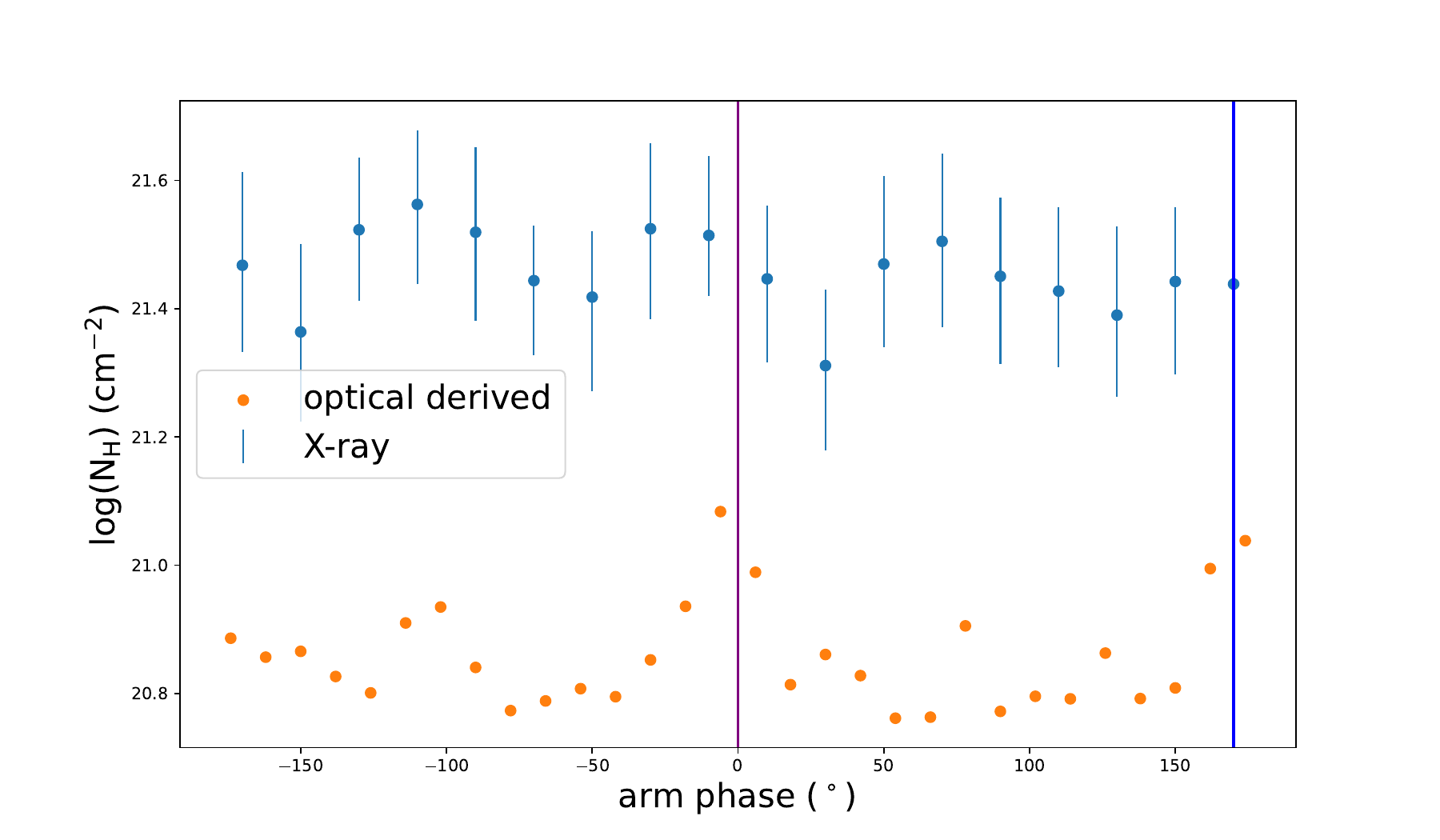}{0.5\textwidth}{}
\fig{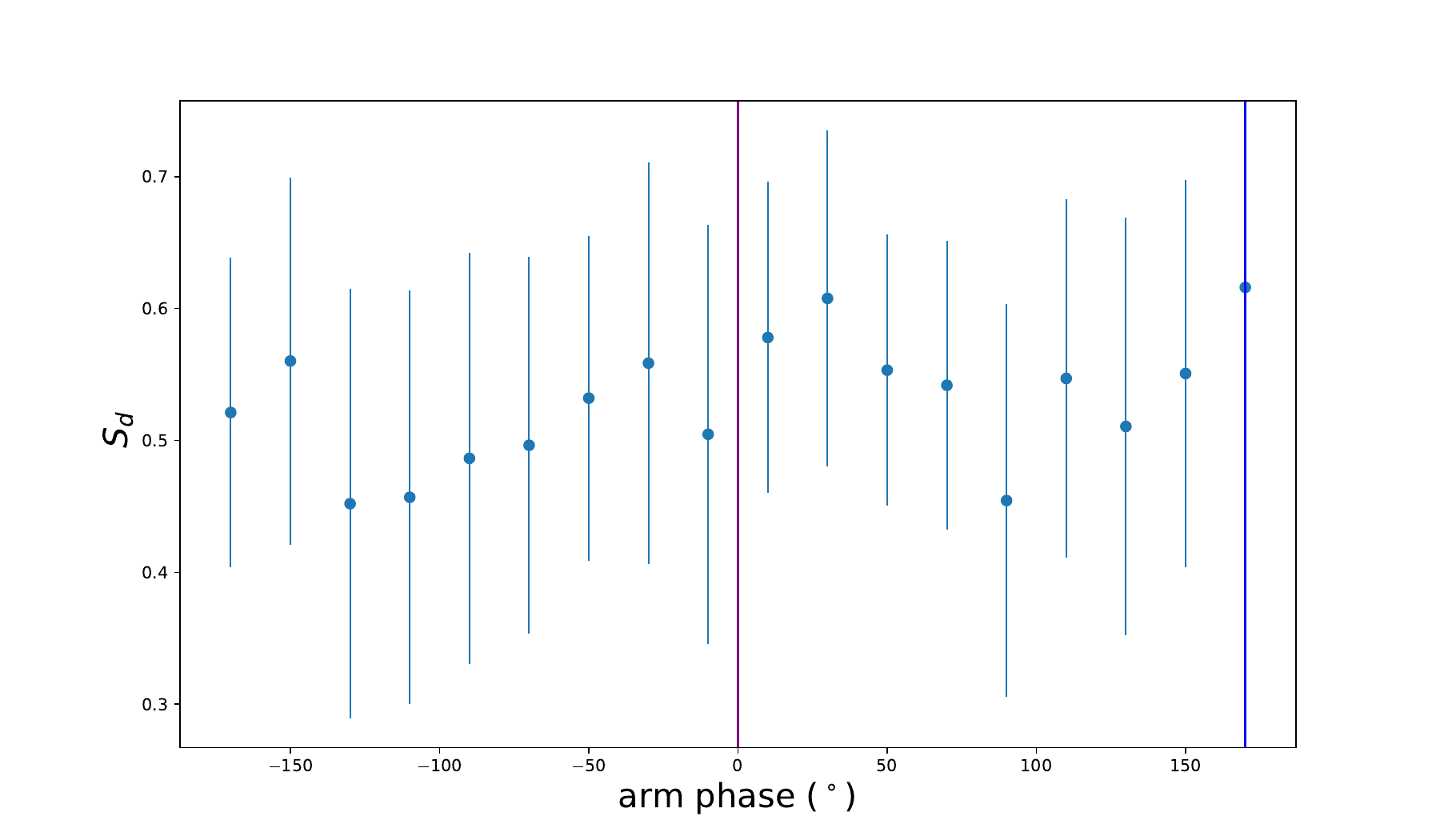}{0.5\textwidth}{}}
\gridline{\fig{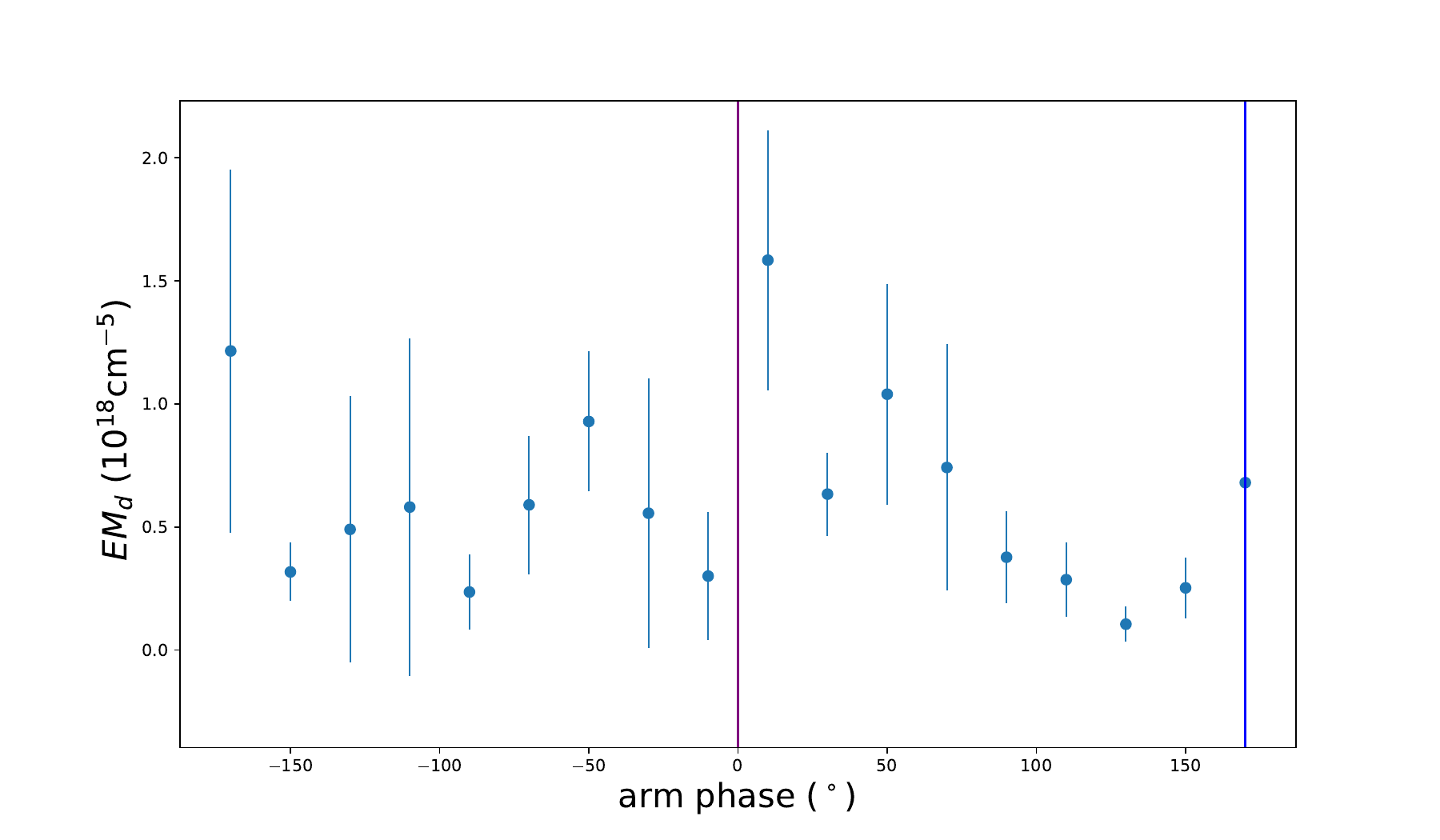}{0.5\textwidth}{}
\fig{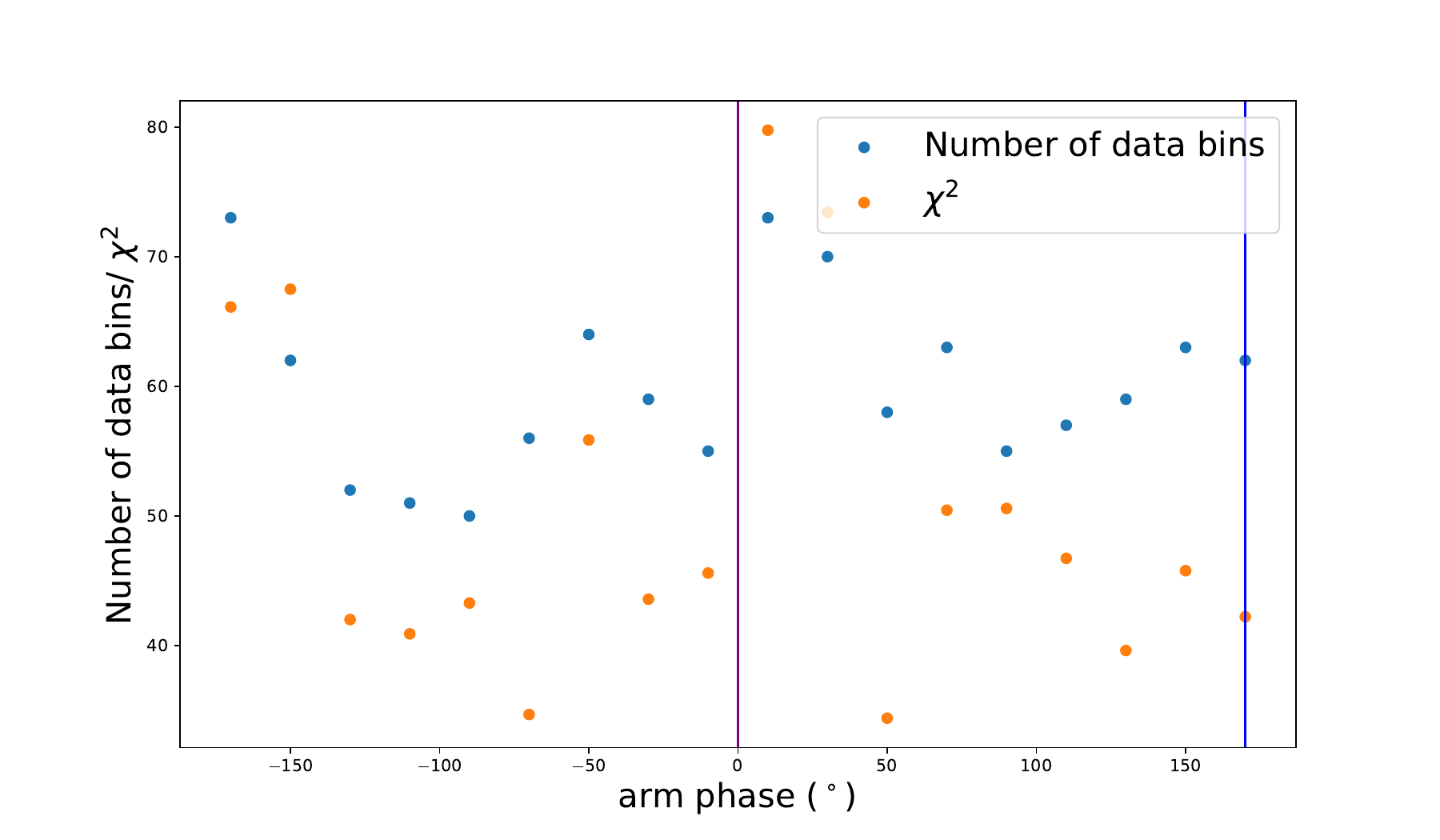}{0.5\textwidth}{}}
\caption{
Local parameter measurements and statistics at different arm phases for Model D: \textbf{Top left panel:} X-ray-absorption-inferred $N_H$ values (blue dots), compared to optical extinction-inferred values (orange);  \textbf{Top right panel:} temperature dispersion of hot plasma in the disk; \textbf{Bottom left panel:} Column emission measure of hot plasma in the disk. \textbf{Bottom right panel:} Number of data bins and $\chi^2$ of each spectrum. The positions of the two spiral arms are marked by the two vertical lines in the same colors as in Fig.~\ref{f:gridding}. The error bars in these panels are not completely independent and somewhat overestimated because of the correlation introduced by the hyperpriors. 
}\label{f:TandnH2}
\end{figure*}

\noindent{\bf Model A}: $(vapec+powerlaw)*tbabs$ or (Disk; see Fig.~\ref{f:geo_diagram} for an illustration of the geometry of the model). We start with this simple reference model, which is typically first used in existing studies of diffuse X-ray emission. Here,  $vapec$ represents all the hot plasma projected in a region, and the power law accounts for unresolved stellar emission. Plasma temperature and $N_H$ are set to have normal distribution priors. When the counting statistics are poor, this model can often provide a simple characterization of the spectral shape of the diffuse X-ray emission. However, the model is clearly inadequate in the present case; it does not give a statistically acceptable fit to the spectra (with a null hypothesis probability less than $10^{-8}$ and gives an unreasonably low metal abundance.  
The observed spectrum appears far too flat to be accounted for by the plasma with a single temperature, plus a simple foreground absorption. However, small dispersions in both temperature and absorption ($\sigma_{x_d}$ and $\sigma_{{\rm log} N_H}$ indicate that the X-ray spectral variation among different regions is small. 

\noindent{\bf Model B}:  $vapec*mxabs_1+powerlaw*mxabs_2*tbabs$ (or Corona/$N_H$mix; Fig.~\ref{f:geo_diagram}; \S~\ref{t:spec}). This model is the same as Model A, except for replacing the absorption in the foreground only with $mxabs$, which is probably a more physical alternative and also has only one parameter $N_H$.  Considering that M51 is an almost face-on galaxy and that the scale height of hot gas is usually significantly higher than the stellar disk, we assume that
$N_H(mxabs_2)=\gamma * N_H(mxabs_1)$, where $\gamma$ is a global parameter determined in our analysis. Apparently, the unresolved stellar emission should also be subject to a foreground absorption of $N_H(tbabs)= [N_H(mxabs_1)-N_H(mxabs_2)]/2$, in addition to $mxabs_2$. 
Our analysis gives $\gamma = 0.96^{+0.04}_{-0.32}$ and shows that this model does not improve the fitting. 
Nevertheless, the comparison between Models A and B suggests that the fitted temperature and metal abundance are sensitive to the assumed relative geometry of the X-ray emission and absorption. The fitted metal abundance of Model B  ($Z_\alpha \sim 1.15$) appears to be much more physical than that of Model A  ($Z_\alpha \sim 0.05$).

However, Model B gives a poorer best fit to the data than Model A, as judged by the statistic $\chi^2/d.o.f.$ in Table~\ref{t:spec}. In either case, the fit is mainly influenced by the high counting statistics of the data near the peak of the observed spectra ($\sim$0.8 keV). For Model A, the temperature is largely determined by the spectral shape on the higher energy side (roughly $\sim$0.8-1 keV) of the peak, while the absorption $TBabs$ with its strong dependence on energy is determined on the lower energy side ($\sim$0.6-0.8 keV). This partly explains the poor fit to the data at $\lesssim 0.5$~keV, in addition to the fact that we have neglected the contribution of CX (see \S~\ref{ss:CX}). For Model B, the absorption $mxabs$ is more distributed, suppressing the plasma contribution more at the peak than Model A for the same N$_H$. This suppression flattens the plasma model spectrum, allowing a better fit at higher energies ($\gtrsim 0.8$~ keV, resulting in a more reasonable metal abundance and a higher temperature. However, the N$_H$ value of $mxabs$, now largely determined by the spectral shape in the $\sim$0.8-1 keV range, leads to an apparent underprediction of the spectral intensity at lower energies and poor fitting statistics in Table~\ref{t:spec}. 

\noindent{\bf Model C} (or Corona+disk; Fig.~\ref{f:geo_diagram}; \S~\ref{t:spec}): 
$ vapec_1*tbabs_1+vapec_2+(vapec_3+powerlaw)*mxabs$.
Here, $vapec_3$ represents the hot plasma, which is evenly mixed with the absorbing gas in the M51 disk, and $vapec_1$ and $vapec_2$ represent the galactic corona, evenly distributed on the far and near sides of the disk, respectively. Only radiation from the far side of the corona is subject to absorption ($tbabs_1$) in the disk with $N_H(tbabs_1)=N_H(mxabs)$ (see Fig.~\ref{f:geo_diagram} for geometry). This helps reduce excess flux at the low-energy end of the X-ray spectrum and the probable overestimation of absorption in Models A and B. The metal abundance of the hot plasma in the disk is allowed to be different from that in the corona and appears to be quite reasonable. The distribution of the corona with its large sound speed is expected to be rather smooth and relatively independent of the spiral arm phase. Therefore, we set $x_c$, $Z_{\alpha,c}$ and $EM_{l,c}$ as global parameters, where $EM_l$ is the column emission measure. It can be simply related to the volume emission measure $EM_V$ and $norm$ by $EM_l=\int n_e n_H dl= \int n_e n_H dV/A =EM_V/A= 10^{14} \times 4\pi D^2 \times norm/A $, where $norm$ is the normalization in the Xspec plasma model $vapec$ or $vlntd$  and $A$ is the area of a region. In contrast, the hot plasma in the disk may vary strongly with the phase. Therefore, we assume that the plasma temperature or $N_H$ of the disk is approximately arm phase dependent. In summary, the model has four local parameters for each region $i$: two local parameters [$x_{d,i}$ and ${\rm log}(N_{H,i})$] with hyperpriors determined by the four hyperparameters [$\mu_{x_d}$, $\sigma_{x_d}$, $\mu_{{\rm log}(N_H)}$, $\sigma_{{\rm log}(N_H)}$] and two ($norm_{d,i}$ and $norm_{po,i}$) without prior; and four global parameters ($Z_{d}$, $Z_{c}$, $x_c$, and $EM_{l,c}$).

This model fits the data well ($\chi^2/dof=987/1006$), and the values of the fitted parameters all appear physically reasonable, except for $Z_\alpha \approx 0.14$, which is somewhat unexpectedly too small. Although local parameters are not tightly constrained due to the limited signal-to-noise ratio (S/N) of individual spectra, the hyper- and global parameters are measured with good accuracy. 
However, a statistically satisfactory fit of the model does not mean that it cannot be improved. Model C assumes two distinct temperatures, which is simple, but is probably hardly physical. The fitted average disk temperature is 0.79 keV, whereas the corona temperature is 0.18 keV.  This large temperature difference, together with the low metal abundance of the corona, appears to be problematic. 

\noindent{\bf Model D:} (log-T Corona+Disk/$N_H$mixed; Fig.~\ref{f:geo_diagram}; \S~\ref{t:spec}): $ vlntd_1*tbabs_1+vlntd_2+(vlntd_3+powerlaw)*mxabs$. This model represents a step towards improving Model C, by replacing the single-temperature plasma ($vapec$) with the lognormal temperature distribution ($vlntd$) for both the disk and the corona.  
Our analysis with this model shows that the observed spectra are mainly contributed by the plasma on the higher temperature side of the lognormal distribution and that  $x$  and $s$ are highly degenerate. Therefore, we choose to constrain $s_{d}$ with a normal hyperprior to characterize the range of the lognormal temperature dispersion. So, similar to Model C, Model D still has four hyperparameters: $\mu_{s_{d}}$, $\sigma_{s_{d}}$, $\mu_{{\rm log}(N_{H})}$, and $\sigma_{{\rm log}(N_{H})}$. In addition, the model has six global parameters: $x_{d}$, $Z_{\alpha,d}$, $x_{c}$, $s_{c}$, $EM_{l,c}$, and $Z_{\alpha,c}$, together with four local parameters for each region $i$: two [$s_{d,i}$ and ${\rm log}(N_{H,i})$] with hyperpriors and two ($norm_{d,i}$ and $norm_{po,i}$) still without a prior. 

Table~\ref{t:spec} and Fig.~\ref{f:spectrum} show that Model D fits the data well, while the corner plot (Fig.~\ref{f:MCMC2}) demonstrates the uncertainties in the estimations of the global parameters. We further use the F-test to compare Models C and D. The F ratio is 24.5, indicating that Model D is significantly better than Model C. The best fit $x_d$ or $x_c$ is small $\sim -1.0$, consistent with the finding in M83 \citep{Wang2021}. The lognormal temperature dispersion of the plasma in the disk ($\mu_{s_d} \sim 0.53$) is also consistent with that measured for M83. Although for the latter galaxy, the disk and corona contributions, as well as their respective absorptions, are not separately modeled, this comparison is still meaningful, since the disk contribution dominates the observed spectrum of a typical region. 

It is interesting to compare the hot plasma properties of the galactic disk and halo. Within the error range, there is no obvious difference in the temperature distribution and metallicity of the disk and corona plasma. By comparing $EM_{l,d}$ with $2EM_{l,c}$ in the bottom panel of Fig. \ref{f:TandnH2}, it can be concluded that the main reason why the spiral arm region is brighter than the inter-arm region is that the emission measure of the disk plasma is higher. This means that it has a higher filling factor or a higher density in the spiral arm region. Considering that the disk here is defined as a plasma mixed with cold gas, the thickness of the disk along the line of sight should be much smaller than that of the corona. Therefore, the fact that $EM_{l,d}$ and $2EM_{l,c}$ (which represent the two sides of the corona) are similar means that the mean density of the disk plasma is at least several times higher than that of the corona plasma.

We show the distribution of the local parameters as a function of the arm phase in Fig.~\ref{f:TandnH2}.  The average temperature or the proportion of high-temperature gas in the arm regions is generally greater than in the interarm regions. However,  there is no clear evidence for an enhanced effective X-ray absorption toward the arm regions.
\begin{figure}
\fig{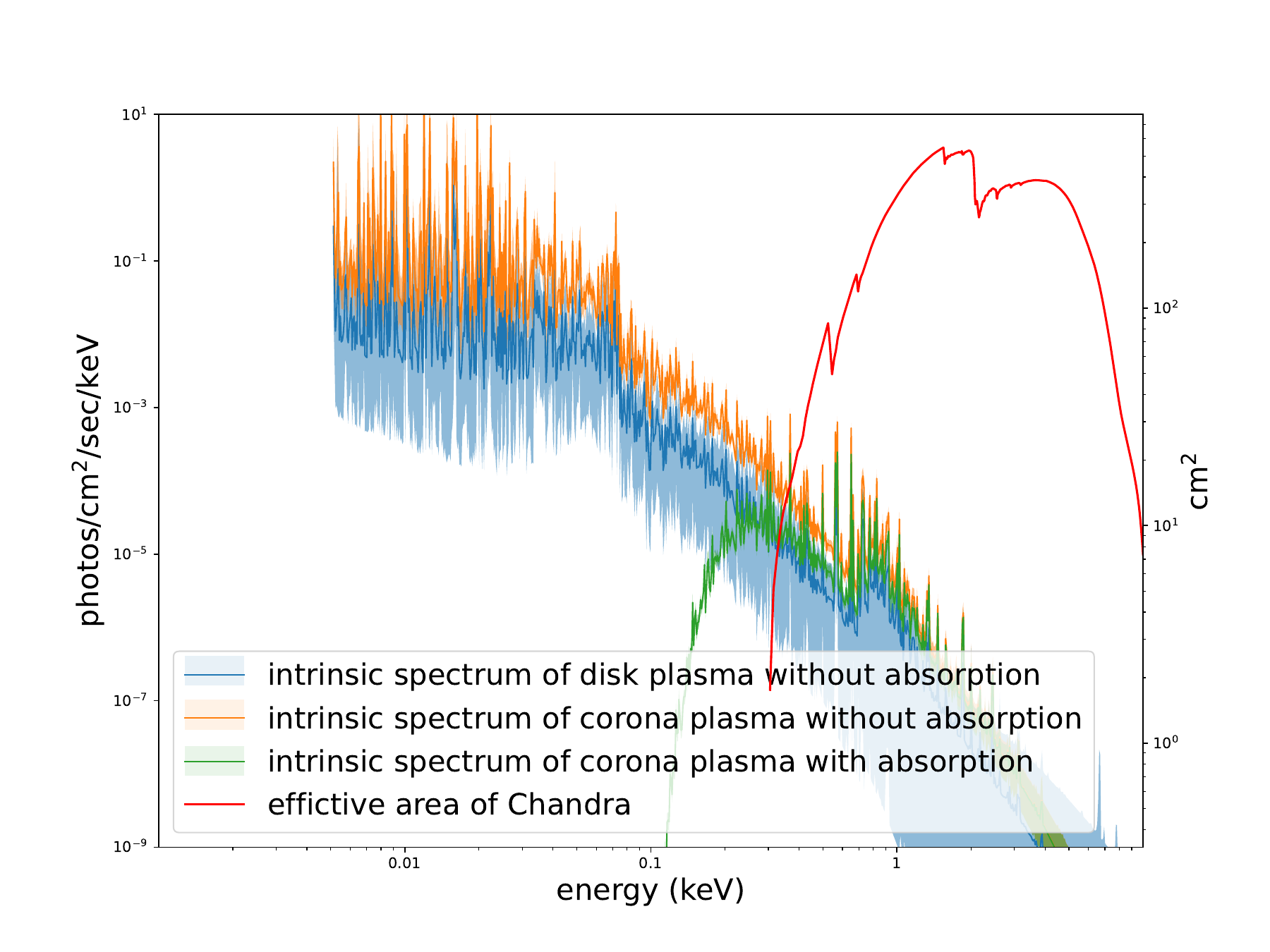}{0.5\textwidth}{}
\caption{Demonstration of the contributions of different components and the X-ray absorption effect (left y-axis), as well as the instrumental effective area (right y-axis). The components are based on the best-fit Model D of an arm phase region. The shadows show the 3$\sigma$ uncertainty range. }\label{f:eff_demo}
\end{figure}

Fig.~\ref{f:spectrum} compares the best fit Model D with two sample spectra of different arm-phase regions, as well as the total integrated spectrum of the annulus considered in the present work. There are statistically significant deviations between the model and the integrated spectrum, largely due to its high signal-to-noise ratio (S/N). Most of them are relatively small ($\lesssim 20-30\%$) and probably within or close to the systemic uncertainties of the calibration and modeling of the data. The exception is the soft X-ray excess below 0.5- 0.6~keV, which reaches about a factor of 2. Such deviations are similar or even more prominent for other spectral models considered here.  This soft X-ray excess and the deviations in general can be most reasonably explained by neglecting the CX contribution (\S~\ref{ss:CX}).  However, the self-consistent inclusion of the CX contribution in our spectral modeling remains difficult. The present CX model depends on the specific temperature of the hot plasma and its velocity relative to that of the cool gas \citep[e.g.,][]{Smith2012,Zhang2014}.  Although our preferred plasma modeling assumes a temperature distribution (e.g., Model D), we demonstrate the contribution, using the latest version of the CX spectral model, ACX \citep[][see also http://www.atomdb.org/CX/]{Smith2012}.  We assume that the plasma undergoing CX in each region can be modeled with APEC with a characteristic temperature to be fitted and with the same metallicity as the hot galactic disk gas in Model D. Following \citet{Zhang2022}, we further approximate the relative velocity to be the sound velocity of the plasma. The best-fit temperatures, 0.05-0.07~keV,  in the regions are slightly lower than the median temperature of Model D, which is reasonable since the CX is expected primarily at the interface between hot and cool gases, where cooling (e.g., via thermal conduction and turbulent mixing) can also be important. The dashed lines in Figure \ref{f:spectrum} show the plasma plus CX contributions.  The CX contribution is dominated by N\texttt{VII} Ly$\alpha$ and N\texttt{VI} He$\alpha$ lines, plus the OVII He$\alpha$ f line,  in the energy 0.4-0.6~keV range, accounting for much of the soft X-ray residual seen above Model D and confirming the importance of the contribution of CX in the soft X-ray end of the X-ray spectra \citep[e.g.,][]{Zhang2022}.

\section{discussion}\label{sec:discussion}
\subsection{Our favored X-ray spectral model}

We have shown that both Models C and D give statistically acceptable fits to the spectra of the diffuse X-ray emission in M51. We regard Model D with a continuous temperature distribution of diffuse hot plasma as more physical. Not only does it fit the spectra of M51 well, but the fitted parameters all appear reasonable and consistent with existing measurements. Consider the abundance of metals as an example. Although the best-fit value for the corona is $\sim 1.4$ solar, which is higher than $\sim 0.97$ solar for the disk, these two abundances are strongly anticorrelated (Fig.~\ref{f:MCMC2}). With this in mind, the X-ray inferred abundances are consistent with the measurements by \citet{Wei2020} and \citet{Croxall2015} based on optical emission lines of HII regions in the galaxy. If we select only those from our X-ray spectral extraction regions, their measured oxygen metal abundance is 0.85$\pm$0.15. A somewhat higher metal abundance in the hot plasma is expected from the chemical enrichment of the stellar feedback in the galaxy. Therefore, we strongly favor Model D. 

The assumed lognormal distribution for the hyperpriors is a minimalist assumption derived from the central limit theorem. When the generation and cooling of hot gas are controlled by many independent multiplicative random processes, such as shock heating, radiative cooling, turbulent mixing, etc., the central limit theorem should apply.  This is also indicated in numerical simulations of the galactic coronae \citep[e.g.,][]{Vijayan2022}.  The distribution has been demonstrated to apply to the spectral characterization of diffuse hot plasma associated with the giant H II region 30 Doradus \citep{Cheng2021} and with the face-on galaxy M83 \citep{Wang2021}. However, we should point out that only the part of the lognormal distribution with $x \gtrsim -1$ is actually constrained by the data, which is limited not only by the effective area of the instrument, but also by the strong soft X-ray absorption of the ISM (Fig.~\ref{f:spectrum}). It is also worthwhile to note that $kT \approx 0.1$~keV  inferred from the best fit $x_d$ or $x_c$ in Model D is the median temperature, not the emission-measure weighted average temperature of the plasma. The latter is 0.2~keV for a lognormal distribution with $x_c= -1.0$ and $s_c=1.2$, for example.

Quantitatively, the median temperature $x_c$ or $x_d \approx 0.1$~keV we obtained is smaller than those temperatures obtained in most existing studies, assuming diffuse radiation spectra\citep[e.g.,][]{Kuntz2010,Anderson2016,Bogdan2017,Zhang2022}. There are two reasons for this difference. First, since most diffuse radiation spectroscopy studies use single or dual temperature models similar to our model B or C instead of a continuous temperature distribution, and $Chandra$ is insensitive photons below 0.3~keV, the dual temperature model usually returns a temperature component around 0.3~keV plus a higher temperature component to compensate for the high-energy residual, and completely ignores the possibility of the existence of possible lower temperature components. Second, most studies adopted an overly simplified geometry for absorption similar to our model A, which underestimates the radiation of low-temperature gases (see Section \ref{subsec:absorption}).

\subsection{X-ray emission/absorption geometric effects}\label{subsec:absorption}

Previous studies of hot plasma in nearby galaxies have commonly adopted an overly simplified geometry that can be problematic. In general, the absorption is assumed to be entirely foreground, with or without including a component internal to such a galaxy. In the latter case, a Galactic foreground $N_{H,G}$ is typically fixed to a value that may be estimated from the 21~cm survey of the Milky Way Galaxy \citep[e.g.,][]{Anderson2016,Bogdan2017}. The former case is similar to our Model A (Fig.~\ref{f:geo_diagram}), in which the column density of the absorbing gas $N_H$ is set as a free parameter \citep[e.g.,][]{Kuntz2010}. Such assumptions about absorption can easily be implemented with a simple or commonly used multiplicative model, such as $tbabs$ available in Xspec, for example.  The resulting spectral model thus appears simple and allows one to focus on constraining the plasma temperature and/or metal abundance, especially when the counting statistics of the data are quite limited. However, the assumption about the geometry of the X-ray-absorbing gas relative to the X-ray emission affects not only the estimate of $N_H$, but also the plasma temperature or its distribution. Taking a 1-T plasma as an example, its temperature is strongly anticorrelated with $N_H$ in a spectral fit. Therefore, a more physical treatment of the absorption geometry can be essential to obtain reasonable spectral parameters of the diffuse hot plasma in and around nearby galaxies. 

Our preferred Model D presents a way to account for the geometry in a reasonably simple and physical way. The model implements the absorption in the disk, so that only the emission from the far side of the galactic corona is affected. With this more realistic modeling, the observed spectrum is contributed by the corona in two parts, with and without absorption in the disk. 

Furthermore, the implementation assumes that the absorption is intermixed with the emission in the galactic disk. Without dealing with its substructure, this simple treatment at least partially accounts for the differential absorption effect caused by the distributions of the emission (whether stellar or diffuse hot plasma) relative to the absorption (by cool ISM). Therefore, the $N_H$ values obtained from our X-ray analysis are always about a factor of $\sim 2-4$  larger than those determined using stellar extinction, which is assumed to be foreground only (Fig.~\ref{f:TandnH2} upper panel).

\subsection{X-ray emission as the stellar feedback tracer}\label{sec:eff}

Diffuse X-ray emission in a star-forming galaxy like M51 is generally believed to be produced by hot plasma heated by stellar mechanical feedback in the form of supernova shocks and fast stellar winds from massive stars. Therefore, it is interesting to see which fraction of the feedback energy ends up in the radiation by the plasma.  To do so, one can estimate the so-called X-ray radiation efficiency or the ratio of the total luminosity of the emission to the feedback energy. Widely reported values of this efficiency include 5\% \citep{Mineo2012} for almost all nearby star-forming galaxies and 1.4\% for highly inclined disk galaxies \citep[e.g.,][]{Li2013}. Therefore, the predominant fraction of the feedback energy is still missing. 

Here, we explore whether or not this missing galactic feedback energy problem is caused by the oversimplification of X-ray spectral modeling. The X-ray radiation efficiency estimate is sensitive to both the assumed spectral model, especially the plasma temperature distribution, and the energy range used to calculate the luminosity.  Both the absorption by the ISM and the small instrument effective area in soft X-ray (e.g., Fig.~\ref{f:eff_demo}) have very much limited the observation of the emission to be above 0.3 keV for external galaxies, yielding little information about their diffuse hot plasma at temperatures $\lesssim 0.1$~keV. The situation is, of course, also not helped by the oversimplified plasma temperature distribution (e.g., 1-, 2-, or 3-T) and absorption geometry (typically foreground only) \citep[e.g.,][ also see our Model A]{Mineo2012,Li2013}, as discussed above.

With our favored Model D, we can now check if the missing feedback energy problem may be resolved. 
We use the gas distribution in M51 provided by \citet{Vieira2023} to infer the star formation density map and estimate the total star formation rate in the extracted spectrum region as 0.3 ${\rm~M_\odot\ yr^{-1}}$. For direct comparison, we follow the assumption of \citet{Mineo2012} that the stellar mechanical energy input rate is dominated by core-collapse supernovae. The rate can be estimated from the SFR: $\sim$ 1 SN per 100 $yr$ per 1 ${\rm~M_\odot \ yr^{-1}}$.And the total energy of each supernova is assumed to be $10^{51} {\rm~erg}$. We then estimate the stellar mechanical energy input rate as $\sim 9 \times 10^{40} {\rm~erg~s^{-1}}$. By comparing this with the X-ray luminosity, the radiation efficiency can be estimated. Assuming the best-fit Model D (Table~\ref{t:spec}) and setting all absorption to zero, we infer the luminosity of the hot plasma emission as $(1.0 \pm 0.4)  \times 10^{40} {\rm~erg~s^{-1}}$ and $(1.5 \pm 0.3) \times 10^{41} {\rm~erg~s^{-1}}$ in the 0.5-2.0~keV and 0.01-100~keV bands, respectively. The 0.5-2.0~keV efficiency is similar to the estimate of \citet{Mineo2012} within its uncertainty.  However, the 0.01-100~keV luminosity, which can be approximated as the total cooling rate of the hot plasma, is a factor of 15 higher; most of the emission occurs in the 0.01-0.3 keV energy range (see also Fig.~\ref{f:eff_demo}), primarily by plasma in the temperature range $\lesssim 0.1$~keV, an estimate allowed by our lognormal temperature distribution modeling. Therefore, the radiation efficiency is $161\% \pm 40\%$ (statistical uncertainty only). We approximate the total radiative energy release of the hot plasma as its intrinsic emission over the 0.01-100 keV range without absorption and find that the corona plasma contributes $(74 \pm 5)\%$, and the remaining quarter comes from the disk plasma. This suggests that M51 may have a corona, which includes outflows from the galactic disk. 
And the uncertainty in the input rate of the feedback energy should be a factor of $\sim 2$. With these considerations, we conclude that the input of the feedback energy can be largely explained by the cooling of the hot plasma in the galaxy.

An important implication of the above conclusion is that most of the stellar feedback energy could be radiated away in the extreme ultraviolet (EUV) to the ultrasoft X-ray range below $\sim 0.3$~keV. This radiation could significantly affect the ionization states of the surrounding ISM and, depending on its effective opacity, potentially the circumgalactic and even the intergalactic medium.

\section{Summary}\label{sec:summary}

We have performed a state-of-the-art analysis of diffuse X-ray emission in the face-on grand spiral galaxy M51. This analysis has the following main ingredients:
\begin{itemize}
\item    Deep (1.3 Ms) $Chandra$ observations provide unprecedented counting statistics of diffuse X-ray emission, with minimal contamination from discrete sources. 
\item A curvilinear coordinate system allows us to conduct a spatially resolved spectral fitting of regions corresponding to different spiral arm phases. The present work focuses on the spectral analysis of 18 such regions located within an annulus of 45\as-110\as\ radii around the galactic center. 
\item The application of the hierarchical Bayesian approach enables a joint spectral fit of these spectra to maximize the constraints on their parameters, both locally and globally.  
\item Spectral models of varying sophistication are tested, from the nominal 1- or 2-T plasma to more physical treatments that account for the important X-ray absorption geometry relative to the emission and the temperature distribution of the hot plasma.  
\end{itemize}

This study has yielded the following main results and conclusions.

\begin{itemize}
\item    The plasma with a lognormal temperature distribution, together with the galactic disk absorption geometry, provides a good simple characterization of the diffuse X-ray spectra, both statistically and physically (see Model D in Table~\ref{t:spec}).  This modeling has only four locally fitting parameters for each spectrum: the residual point source contribution (assuming a known power law), the X-ray-absorbing gas column density, the dispersion ($s_d$), and normalization of the lognormal distribution of the galactic disk plasma. In the fit to each spectrum, $s_{d}$ for the galactic disk is allowed to vary with a normal hyperprior with its parameters ($\mu_{s_d}$ and $\sigma_{s_d}$) fitted in the joint analysis.  In addition, the lognormal temperature mean ($x_{d}$) and the metal abundance ($Z_{\alpha,d}$) of the galactic disk hot plasma,  as well as all parameters of the galactic corona ($x_c$, $s_{c}$, $norm_{c}$, and $Z_{\alpha,c}$) are jointly fitted across the regions. 
\item The fitted parameters of the spectral model are physically reasonable (Table~\ref{t:spec}). The median temperature/$Z_\alpha$ of the hot plasma is about $10^{-1.00}$~keV/1.0 solar and $10^{-1.07}$~keV/1.4 solar in the galactic disk and corona, respectively. The hyperparameters $\mu_{s_d} \approx 0.53$ and $\sigma_{s_d} \approx 0.10$ characterize the variation of the temperature distribution between different arm phases. The X-ray emission from diffuse hot plasma in spiral arm regions is generally brighter and harder than in inter-arm regions. This phenomenon can be naturally explained by the difference in the emission measure and the broadening of the temperature distribution of these regions. The globally fitted $s_{c} \approx 1.25$ indicates a large temperature dispersion in the corona.
\item The stellar feedback energy could dissipate largely through radiative cooling of the plasma. Prior studies have indicated that diffuse X-ray emission is inefficient, with a luminosity substantially lower than the energy input from stellar feedback. If Model D is reasonably accurate, the missing galactic feedback can be explained by the majority of stellar feedback energy being radiated in the EUV and ultra-soft X-ray ($\lesssim 0.3$~keV) range. Although much of this radiation is absorbed by the ISM, some, particularly from the galactic corona, can escape, ionizing the circumgalactic or intergalactic medium.
\item The mean X-ray-absorbing gas column density is ${\rm log}(N_H) \approx 21.5$, with a dispersion of $0.11$ between different regions of the arm phase. This value is a factor of about 2-4  higher than those inferred from the optical band, due to our adjustment for the X-ray-absorbing gas geometry relative to the corona.

\end{itemize}


\section*{Data Availability}
This paper employs a list of Chandra datasets, obtained by the Chandra X-ray Observatory. 

\bibliography{paper_M51}{}
\end{document}